\documentclass[amsmath, amssymb, aps, prb, floatfix, superscriptaddress,
preprint]{revtex4}
\usepackage{graphicx}% Include figure files
\usepackage{dcolumn}% Align table columns on decimal point
\usepackage{bm}% bold math
\usepackage{amsmath}
\begin{document}

\title{ Anomalous  energy  diffusion in two-dimensional nonlinear lattices}

\author{Jian Wang}
\email{phcwj@hotmail.com}
\affiliation{College of Physical Science and Technology, Yangzhou University, Yangzhou 225002, P. R. China }

\author{Tian-xing Liu}
\affiliation{College of Physical Science and Technology, Yangzhou University, Yangzhou 225002, P. R. China }

\author{Xiao-zhi Luo}
\affiliation{College of Physical Science and Technology, Yangzhou University, Yangzhou 225002, P. R. China }

\author{Xiu-Lian Xu}
\affiliation{College of Physical Science and Technology, Yangzhou University, Yangzhou 225002, P. R. China }

\author{Nianbei Li}
\email{nbli@hqu.edu.cn}
\affiliation{Institute of Systems Science and Department of Physics, College of Information Science and Engineering, Huaqiao University, Xiamen, 361021, P. R. China }

\date{\today}

\begin{abstract}
   Heat transport in one-dimensional (1D) momentum-conserving lattices is generally assumed to be anomalous, thus yielding a power-law divergence of thermal conductivity with system length. However, whether heat transport in two-dimensional (2D) system is anomalous or not is still on debate because of the difficulties involved in experimental measurements or due to the insufficiently large simulation size. Here, we simulate energy and momentum diffusion in the 2D nonlinear lattices using the method of fluctuation correlation functions. Our simulations confirm that energy diffusion in the 2D momentum-conserving lattices is anomalous and can be well described by the L\'{e}vy-stable distribution. We also find that the disappear of side peaks of heat mode may suggest a weak coupling between heat mode and sound mode in the 2D nonlinear system. It is also observed that the harmonic interactions in the 2D nonlinear lattices can accelerate the energy diffusion. Contrary to the hypothesis of 1D system, we clarify that anomalous heat transport in the 2D momentum-conserving system cannot be corroborated by the momentum superdiffusion any more. Moreover, as is expected, lattices with a nonlinear on-site potential exhibit normal energy diffusion, independent of the dimension. Our findings offer some valuable insights into the mechanism of thermal transport in 2D system.

\end{abstract}
%\pacs{}

%\keywords{}=

\maketitle
\newpage

\section{Introduction}

The conventional Fourier's law for heat conduction postulates\cite{mathp2000} a proportional relation between heat fluxes $J$ and  temperature $T$ gradient  as $J=-\kappa \nabla T$, where thermal conductivity $\kappa$ is independent of system size and is a constant at given temperature. The Fourier's law has witnessed a good validation in the three-dimensional (3D) bulk systems.  However, the breakdown of Fourier's law has recently been confirmed\cite{mathp2000,Lepri2003c,Dhar2008,Wang2008a,review2016,Li2012a,Gu2018a} in low-dimensional systems. Based on the extensive numerical simulations and various analytical approaches as well as experimental results, it is generally believed that heat conduction in low-dimensional systems does not observe Fourier's law any more. The understanding of heat conduction in low-dimensional system\cite{mathp2000,Lepri2003c,Dhar2008,Wang2008a,review2016,Li2012a,Gu2018a}  becomes  a challenging  problem both for statistical physics\cite{mathp2000,Lepri2003c,Dhar2008,Wang2008a,review2016} and applied low-dimensional thermal engineering\cite{Li2012a,Gu2018a}.

 For one-dimensional (1D) system, thermal conductivity $\kappa$ in the 1D Fermi-Pasta-Ulam $\beta$($\text{FPU-}\beta$) nonlinear lattices was first found\cite{Lepri1997} to diverge with the system size $N$ as $\kappa \varpropto N^{\alpha}$ with $0<\alpha<1$. The size-dependent thermal conductivity  breaks the prediction of Fourier's law, implying that thermal conductivity $\kappa$ is no longer an intrinsic property of a material. This non-fourier heat transport is referred to as anomalous heat transport. Apart from the subsequent investigations\cite{Lepri1998,PhysRevE.57.2992,Lepri2005,Mai2007} on the $\text{FPU}$ chains,  similar length-dependent thermal conductivity was also observed in the diatomic Toda lattice\cite{Hatano1999,Grassberger2002,xiong2017,Li2018} and random collision model\cite{Deutsch2003}. Although much effort such as mode-coupling theory\cite{Lepri1998a}, hydrodynamical theory\cite{Narayan2002b} and nonlinear fluctuating hydrodynamics\cite{Beijeren2012,Spohn2014} has been devoted to the universal exponents $\alpha$, which can be largely independent of the microscopic details, there is still on debate about the classification of anomalous behavior\cite{review2016} in terms of universality classes.  In the above one-dimensional nonlinear lattices, it should be noted the total momentum  is conserving. On the other hand, normal heat conduction observing Fourier's law has been demonstrated in the 1D Frenkel-Kontorova lattices\cite{PhysRevE.57.2992,Tsironis1999} and $\phi ^4$ lattices\cite{Hu2000a,Aoki2000}, where the total momentum is not conserving due to the presence of external on-site potentials. Thus, an intuitive conjecture can be made that momentum conservation in low-dimensional lattices leads to\cite{Prosen2000} anomalous heat conduction. However, a contradictory result is founded in the 1D coupled rotator lattices\cite{Giardina2000} which reveal normal heat conduction in spite of momentum conservation. In other words, the conservation of momentum  does not ensure anomalous heat conduction in low-dimensional lattices. To identify the origin of anomalous heat transport, a further hypothesis\cite{Li2015} that normal(anomalous) spread of excess momentum density gives rise to normal(anomalous) heat conduction was made for 1D nonlinear lattices.

 Furthermore, understanding thermal transport in 2D systems is not only for completeness of the theoretical framework\cite{mathp2000,Lepri2003c,Dhar2008,review2016}, but is also of great interest for the possible technological applications\cite{review2016,Li2012a,Gu2018a} in 2D realistic materials like graphene\cite{Nika2017}. Theoretical investigations in the 2D  nonlinear lattices are less developed on account of the technical difficulties which are resulting from the increased dimension. Based on the mode-coupling theory\cite{Lepri2003c,Dhar2008,review2016}, it has been conjectured that thermal conductivity in the 2D nonlinear lattices with the conserving momentum will diverge logarithmically with system size. Numerical simulations in the 2D $\text{FPU-}\beta$ nonlinear lattices\cite{Lippi2000a} and in the disk lattices\cite{PhysRevE.82.030101} with vector displacements have verified such logarithmical divergences. In contrast, a power-law divergence of thermal conductivity with system size is observed\cite{Shiba2008a} in  the 2D $\text{FPU-}\beta$ nonlinear lattices. A recent study with scalar displacements\cite{Wang2012a} reveals a power-law divergence in the 2D $\text{FPU-}\beta$ nonlinear lattices and a logarithmically divergent thermal conductivity for the purely quartic lattices. Besides, normal heat conductivity was observed\cite{Savin2016} in  the 2D scalar lattices. A possible explanation for these differences in the simulation results maybe lies in the strong finite-size effects\cite{Wang2012a} within affordable computational resources.

  Heat transport in 2D systems is therefore far from being clear.  Because of the rapid increase of particle numbers in the larger 2D system, it is hard to numerically calculate heat conduction via either the direct simulations of nonequilibrium molecular dynamics or the equilibrium Green-Kubo approach. Microscopically, heat conduction can in principle be related to heat or energy diffusion process. For example, Fourier's law is characterized by the normal spread of energy carriers while anomalous heat conduction in low-dimensional systems implies\cite{Lepri2003c,Dhar2008,Liu2014b,review2016} anomalous energy diffusion. The nonlinear hydrodynamic theory \cite{Beijeren2012,Spohn2014} analytically describes heat diffusion process through the scaling forms of coupled correlation functions for the stretch, momentum and energy, respectively, and has significantly contributed to understanding heat diffusion in 1D systems.  However, it is difficult to apply the analytical nonlinear hydrodynamic theory to 2D systems. Energy diffusion in 2D nonlinear lattices acts as another important approach to understanding heat transport in 2D systems, since the diffusion method can circumvent the finite-size problem confronted by the direct simulation of heat conduction. Nevertheless, the approach of energy diffusion is difficult to be utilized in 2D nonlinear lattices on account of heavy computations. To overcome this obstacle of heavy computations, our present work has successfully accelerated the computation through graphics processing unit\cite{cuda} (GPU). The aim of this paper is to investigate energy and momentum diffusion in the three types of 2D nonlinear lattices, thus elucidating the corresponding characteristics of heat transport in 2D systems.

 The paper is organized as follows. In Sec.II, we first introduce three typical models of 2D nonlinear lattices and then describe the adopted simulation approach using the energy/momentum fluctuation correlation function. In Sec.III, we present our main results of energy and momentum diffusions for the three 2D nonlinear lattices, accompanied by a discussion on the possible mechanism in terms of  L\'{e}vy walk distributions. Finally, conclusions are drawn in Sec.IV.

\section{Models and Methods}
\subsection{Models}

Most majority of the studies on energy diffusion in nonlinear lattices\cite{H.Zhao2006,Chen2013,Li2015,Gao2016c,xiong2017,Xiong2018a,Xiong2018b} has been devoted to 1D systems. So far, no investigation of energy diffusion or of momentum diffusion has been made in 2D nonlinear lattices. Only a few work on heat transport in 2D nonlinear lattices was performed\cite{Lippi2000a,Yang2006a,Shiba2008a,PhysRevE.82.030101,Wang2012a,Savin2016} through the direct nonequilibrium molecular dynamics simulations or the equilibrium Green-Kubo approach.  Here we consider energy diffusion in the 2D square lattices made of $N_x\times N_y$ atoms with a vector displacement of $\textbf{q}_{i,j}$. $N_x$ and $N_y$ denote the number of atoms in the $x$ and $y$ direction, respectively. The equilibrium positions of the atoms coincide with the lattice sites, labelled by a pair of integer indices $\{(i, j), {i=1, N_x; j=1, N_y}\}$. For simplification, each atom in those lattices only interacts with its nearest neighbors that characterize the short-range interatomic forces in real solids. The general Hamiltonian for the adopted 2D  nonlinear lattices can be written as
\begin{equation}
\label{Hamiltonian for 2d lattices}
H=\sum_{i=1}^{N_y}\sum_{j=1}^{N_y}\big[\frac{\textbf{p}_{i,j}^{2}}{2m}+V(|\textbf{q}_{i+1,j}-\textbf{q}_{i,j}|)+V(|\textbf{q}_{i,j+1}-\textbf{q}_{i,j}|)+U(\textbf{q}_{i,j})\big],
\end{equation}
where $\textbf{q}_{i,j}$ is the vector displacement from its equilibrium position of the atom on the $(i,j)$ lattice site and $\textbf{p}_{i,j}$ corresponds to its momentum vector. $m$ is the atom mass and has been set $m=1$ without loss of of generality. The interaction potential is taken as $V(r)=\frac{1}{2}kr^{2}+\frac{1}{4}\beta r^{4}$. The term $U=\frac{1}{4}gr^{4}$ denotes the on-site potential, which breaks the momentum conservation. To compare our results of energy diffusion with previous simulations\cite{Lippi2000a,Yang2006a,Shiba2008a,Wang2012a,Savin2016} of heat conduction, we consider three typical types of 2D nonlinear lattices: the $\text{FPU-}\beta$ model  with $k=1,\beta=1, g=0$; the purely quartic lattice with $k=0,\beta=1, g=0$  and the $\phi^4$ model with $k=1,\beta=0, g=1$. The $\text{FPU-}\beta$ model has been widely used\cite{Lippi2000a,Yang2006a,Shiba2008a,Wang2012a} for studying nonlinear behaviors and heat transport in low-dimensional nonlinear lattices. The purely quartic lattice can be regarded as\cite{Wang2012a,review2016} the high temperature limit of the $\text{FPU-}\beta$ lattice. The $\phi^4$ model\cite{Hu2000a,Aoki2000,phi4book} without the momentum conservation is studied to demonstrate the effect of the momentum conservation on energy diffusion in 2D nonlinear lattices. To show the dimensional crossover of energy diffusion in the 2D nonlinear lattices, the length $N_x$ is fixed to be 1023, while the lattice width $N_y$  varies from 1(1D) to 1024(2D). Consequently, the largest number of particles in the nonlinear lattices is up to 1047552 during our simulations.

\subsection{Methods}

To make a direct comparison with the hydrodynamics theory\cite{Beijeren2012,Spohn2014}, we focus on energy
and momentum diffusion in the above 2D nonlinear lattices. The nonlinear
hydrodynamic fluctuation theory
states\cite{Beijeren2012,Spohn2014,Das2014a} that anomalous
energy/momentum  diffusion in nonlinear lattices can be characterized by the
scaling forms of space-time correlation of fluctuation functions. Here we define
the spatiotemporal correlation of fluctuation function\cite{H.Zhao2006,Chen2013} of
energy $\rho_{E}(i,t)$ and momentum $\rho_{P}(i,t)$ fluctuation for a microcanonical
system  as
\begin{subequations}
\label{correlation}
\begin{equation}
 \rho_{E}(i,t)  =
\frac{\big\langle\Delta{H}_{i}(t)\Delta{H}_{0}(0)\big>}{
\big\langle\Delta{H}_{0}(0)\Delta{H}_{0}(0)\big>}+\frac{1}{N_{x}-1},  \\
\end{equation}
\begin{equation}
  \rho_{P}(i,t)  =
\frac{\big\langle\Delta{P}_{i}(t)\Delta{P}_{0}(0)\big>}{
\big\langle\Delta{P}_{0}(0)\Delta{P}_{0}(0)\big>}+\frac{1}{N_{x}-1}.
\end{equation}
\end{subequations}
The coarse-grained local energy density $H_{i}(t)$
and momentum density $P_{i}(t)$  are both summed over the energy of the
$i$th column of atoms in 2D nonlinear lattices. The energy/momentum
fluctuation corresponds  to
$\Delta H_{i}(t)=H_{i}(t)-\bar{H}_{i}$ and
$\Delta{P}_{i}(t)=P_{i}(t)-\bar{P}_{i}$,  respectively.
The spatiotemporal average $\langle\cdot\rangle$ in Eq.(\ref{correlation}) is performed
along the $x$ direction and with time because of the
spatial and time translational invariance in equilibrium states. The notation $\bar{H}_{i}$ and $\bar{P}_{i}$ denote the averaged local density of energy and momentum, respectively. The length $N_x$ is
the number of lattice sites along the $x$ direction. From the perspective
of hydrodynamics theory, the spatiotemporal
correlation function $\rho_{E}(i,t)$ and $\rho_{P}(i,t)$ can be viewed as the fingerprint
for the behaviors of energy and  momentum diffusion, corresponding to the
correlation of heat modes and sound modes, respectively. To quantify the overall energy
diffusions on lattices, we also have calculated the mean-square deviation
(MSD) of energy $\langle\Delta x^{2}(t)\rangle_{E}$ as
\begin{equation}
\label{msd}
 \langle\Delta x^{2}(t)\rangle_{E} =\sum_{i}i^{2}\rho_{E}(i,t).
\end{equation}

During the  numerical simulations, periodic boundary conditions are applied  in  the both $x$
 and $y$ directions.  The energy and momentum fluctuation correlation function are calculated in the equilibrium state at temperature $T=0.5$ for purely quartic lattice and the $\text{FPU-}\beta$ model, and $T=3.0$ for the $\phi^4$ model. Here, the temperature has been chosen to be high enough so that the nonlinear interactions are excited during the simulations.  The number of the ensemble-averaging in the spatiotemporal correlation function of Eq.(\ref{correlation}) is up to $10^8$  after an equilibrium state is firstly prepared from properly assigned random states. The velocity-Verlet algorithm is adopted for integrating the motion equations. Owing to the heavy computations from the large system of 2D lattices, we employ the graphics processing unit(GPU) to implement the parallel acceleration.

\section{Results and Discussions}

We probe the diffusion behavior of energy and momentum by simulating the spatiotemporal fluctuation correlations in three typical nonlinear lattices: the purely quartic lattices, the $\text{FPU-}\beta$ lattices and the $\phi^4$ model. The purely quartic lattices and the $\text{FPU-}\beta$ lattices hold the momentum conservation while the $\phi^4$ model breaks the conservation of momentum owing to the onsite potential.  To demonstrate the dimensional crossover of energy diffusion from 1D to 2D, we increase the lattice width $N_y$, varying from 1 to 1024 with a fixed length $N_x=1023$.  We first present energy diffusion in three types nonlinear lattices, elucidating the nature of heat mode in 2D nonlinear lattices. Next, we investigate the diffusion of momentum, in order to understand the dimensional-crossover features of sound modes.

\subsection{Profiles of energy fluctuation correlation function}

  \begin{figure}[hb]
\includegraphics[width=0.7\columnwidth]{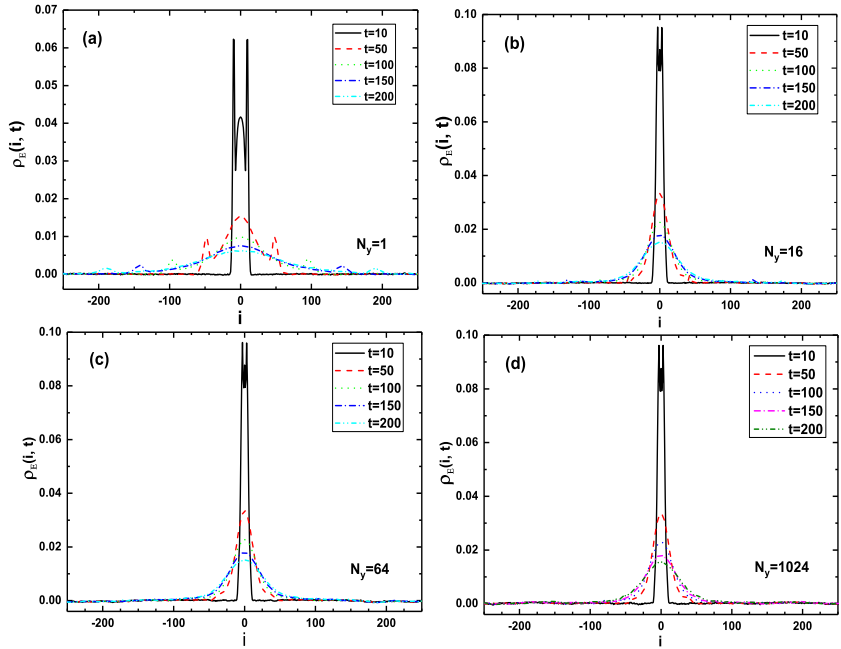}
\caption{\label{fig1:Equartic_diff} (Color online). The spatial profiles of the energy fluctuation correlation function $\rho_{E}(i,t)$ for the purely quartic lattices with increasing width (a) $N_y=1$, i.e. 1D lattices, (b) $N_y=16$, (c) $N_y=64$, (d) $N_y=1024$. The spatial profiles of $\rho_{E}(i,t)$ at times $ t=10,50,100,150, \text{and} \; 200 $ are labelled by the different lines as indicated in the figure.    }
\end{figure}

\begin{figure}[htb]
\includegraphics[width=0.7 \columnwidth]{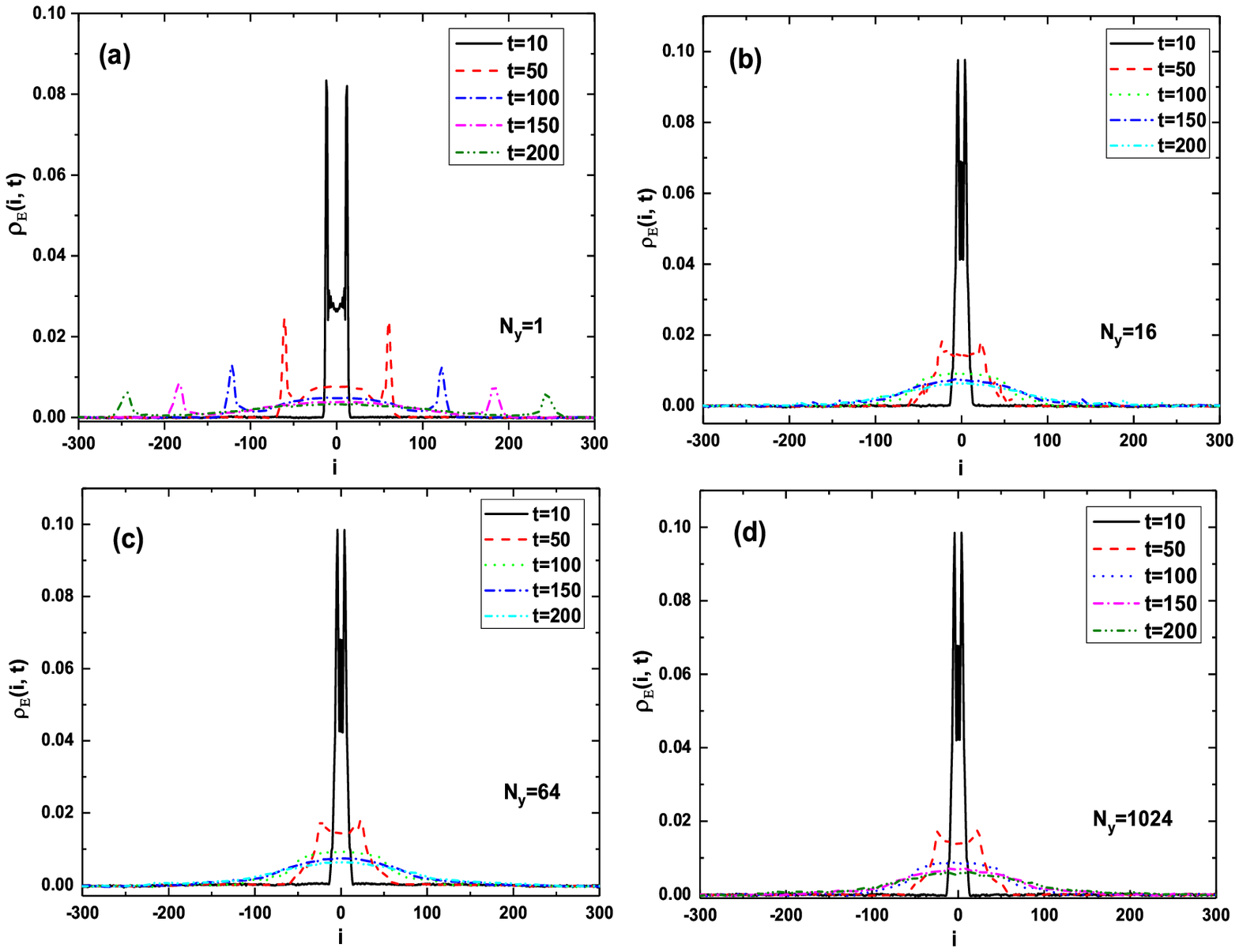}
\caption{\label{fig2:Ebeta_diff} (Color online).  The spatial profiles of the energy fluctuation correlation function $\rho_{E}(i,t)$  for the $\text{FPU-}\beta$ lattices with increasing width (a) $N_y=1$, i.e. 1D lattices, (b) $N_y=16$, (c) $N_y=64$, (d) $N_y=1024$. The spatial profiles of $\rho_{E}(i,t)$  at times $ t=10,50,100,150, \text{and} \; 200 $ are labelled by the different lines shown in the figure. }
\end{figure}

\begin{figure}[htb]
\includegraphics[width=0.7 \columnwidth]{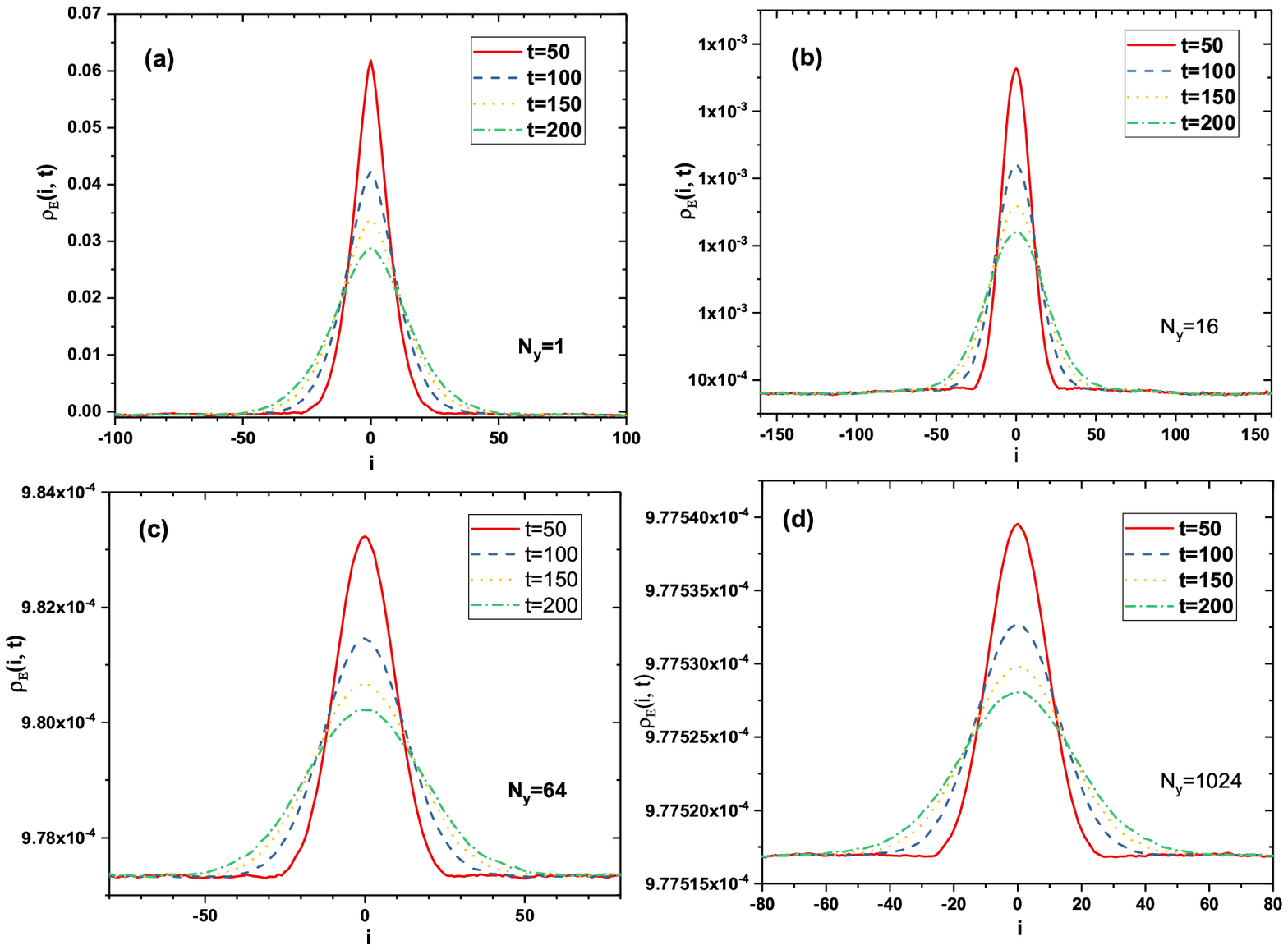}
\caption{\label{fig3:Efei4_diff} (Color online).  The spatial profiles of the energy fluctuation correlation function $\rho_{E}(i,t)$  for the $\phi^4$ lattices with increasing width (a) $N_y=1$, i.e. 1D lattices, (b) $N_y=16$, (c) $N_y=64$, (d) $N_y=1024$. The spatial profiles of $\rho_{E}(i,t)$  at times $ t=50,100,150, \text{and} \; 200 $ are labelled by the different lines illustrated in the figure. }
\end{figure}

The spatial profiles of energy fluctuation correlation function $\rho_{E}(i,t)$ with varying width for the purely quartic lattices, the $\text{FPU-}\beta$ lattices and  the $\phi^4$ lattices are depicted in Fig.~\ref{fig1:Equartic_diff}, Fig.~\ref{fig2:Ebeta_diff} and Fig.~\ref{fig3:Efei4_diff}, respectively. As the probability distribution function, $\rho_{E}(i,t)$ exhibits the nonnegative property in the figure.  As is shown in the figures, there is a prominent central peak in each spatial profile of $\rho_{E}(i,t)$, which corresponds to the heat mode according to the hydrodynamical theory\cite{Beijeren2012,Spohn2014,Das2014a}.  For the 1D nonlinear lattices i.e. $N_y=1$, our calculated  $\rho_{E}(i,t)$ is qualitatively consistent with the previous results\cite{H.Zhao2006,Chen2013,Li2015,Gao2016c,xiong2017,Xiong2018a,Xiong2018b,phi4book}. In particular, on the 1D quartic and $\text{FPU-}\beta$ lattices, there are two observable right/left-moving side peaks along both sides of the central heat mode. These side peaks originate from a strong coupling\cite{Das2014a} between the heat modes and the sound modes by the mode cascade theory\cite{Lee-Dadswell2005}. Thus, it can be inferred from the prominent side peaks that the energy and momentum transport is strongly coupled in the momentum-conserving 1D nonlinear lattices. This strong coupling between heat and sound modes  may contribute to anomalous energy transport by the mode cascade theory\cite{Lee-Dadswell2005}. In comparison with Fig.~\ref{fig1:Equartic_diff}(a), the side peaks of the 1D $\text{FPU-}\beta$ lattices in Fig.~\ref{fig2:Ebeta_diff}(a) are much stronger and decay more slowly. We believe the salient feature of side peaks in the $\text{FPU-}\beta$ model may result from the harmonic interactions, which not only give rise to the acoustic phonon modes but also strongly enhance the coupling between the heat and sound modes.  In contrast, it can be seen from Fig.~\ref{fig1:Equartic_diff} and Fig.~\ref{fig2:Ebeta_diff} that these side peaks become rapidly diminished with the increase of lattice width for both the purely quartic lattices and the $\text{FPU-}\beta$ lattices. As illustrated in  Fig.~\ref{fig1:Equartic_diff} and Fig.~\ref{fig2:Ebeta_diff}, we can find that the  side peaks almost vanish  away when the width of lattices $N_y\geq16$. Such disappearance of side peaks for the wider lattices may suggest that the coupling between heat and sound modes becomes smaller with the increase of lattice width, possibly implying a subdued heat transport in the momentum-conserving 2D nonlinear lattices.  On the other hand, for the $\phi^4$ lattices, there are no side peaks in Fig.~\ref{fig3:Efei4_diff}, independent of the lattice width. This observation of the $\phi^4$ lattice with the on-site potential is in agreement with the previously reported spatial profiles of energy fluctuation correlation function\cite{H.Zhao2006,phi4book} for the 1D $\phi^4$  model. We think that the side peak generally does not occur for the lattices without the conserving momentum, where heat transport may be carried out mainly in terms of heat diffusion.   Apart from the significant change of side peaks with the lattice width, the prominent cental peak of heat modes in the spatial profiles of Fig.~\ref{fig1:Equartic_diff}, Fig.~\ref{fig2:Ebeta_diff} and Fig.~\ref{fig3:Efei4_diff} dominates the feature of energy diffusion in low-dimensional lattices. Next, we characterize the features of these central peaks based on the L\'{e}vy walk theory using the scaling analysis.

To investigate the probability character of the central peak of heat mode, we first turn to the predictions of 1D nonlinear lattice by the recent nonlinear fluctuating hydrodynamic theory\cite{Beijeren2012,Spohn2014,Das2014a}, which relates its heat and sound modes to the fluctuation correlations of three conserving quantities: energy, momentum and stretch. Generally, this spatial-temporal fluctuation correlation function such as $\rho_{E}(i,t)$ observes a scaling invariant relationship\cite{Zaburdaev2015} as
\begin{equation}
\label{scaling}
 t^{\frac{1}{\gamma}}\rho_{E}(i,t')\simeq \rho_{E}(i/t^{\frac{1}{\gamma}},t),
\end{equation}
 with the scaling exponent $1\leq \gamma\leq 2$ . Different behaviors of diffusion can be characterized\cite{firststeprw} by the exponent $\gamma$: ballistic diffusion $\gamma= 1$, superdiffusive diffusion $1<\gamma<2$ and normal diffusion(i.e. Gaussian distribution) $\gamma= 2$. Phenomenologically, the dynamics of diffusion process can be modelled\cite{firststeprw} by the L\'{e}vy walk using the L\'{e}vy-stable distribution $f_{LW}^{\gamma}(m,t)$ that is the Laplace-Fourier transform of L\'{e}vy characteristic function $e^{-\mid k\mid^{\gamma}t}$. To put it more straightforwardly, the diffusion property of the fluctuation correlation function such as $\rho_{E}(i,t)$ is determined by the scaling exponent $\gamma$ and  has the same mathematical property as the L\'{e}vy-stable distribution $f_{LW}^{\gamma}(i,t)$. Anomalous energy diffusion in our simulations can be identified through the superdiffusive diffusion $1<\gamma<2$. For the conserving quantity, the scaling exponent $\gamma$ can be extracted\cite{H.Zhao2006,Li2015} from the height of the central peak in the fluctuation correlation function. Now, we start to clarify the diffusion property of heat mode using the above approach.
 \begin{figure}[ht]
\includegraphics[width=0.8 \columnwidth]{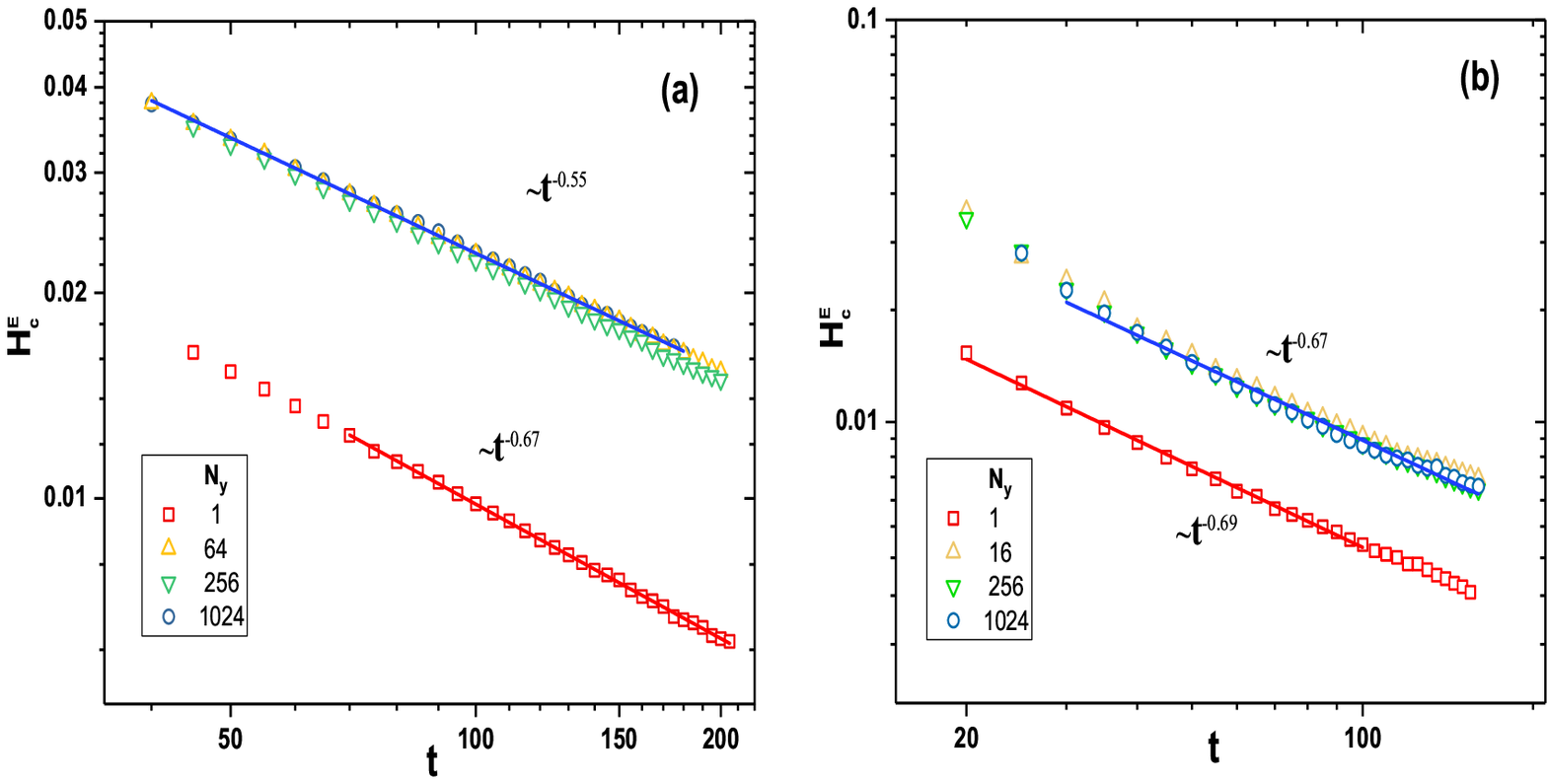}
\caption{\label{fig4:Eh_decay} (Color online). The decay of the peak height of the energy fluctuation correlation function $H_{C}^{E}$ as a function of time (a) for the  purely quartic lattice, (b) fort the $\text{FPU-}\beta$ lattices. The decay of height is fitted to a power-law distribution $H_{C}^{E} \sim t^{-1/\gamma}$ as shown in the figure.  }
\end{figure}
 For the purely quartic lattices and the $\text{FPU-}\beta$ model with the conserving momentum, we  extract the scaling exponent $\gamma$ from the decay of peak height $H_c^E$ of the central heat mode. The log-log plots of the peak height $H_c^E$  as a function of time are depicted in Fig.~\ref{fig4:Eh_decay}, respectively. As can be seen from the figure, a power-law decay $H_c^E \sim t^{-1/\gamma}$ can be explicitly fitted. The rate of decay $1/\gamma$ of $H_c^E$ for the purely quartic lattices decreases from $1/\gamma=0.67$ for 1D($N_y=1$) to $1/\gamma=0.55$ for 2D ($N_y=1024$) as illustrated in Fig.~\ref{fig4:Eh_decay}(a). In contrast, the decay rate $1/\gamma$ of the central peak for the $\text{FPU-}\beta$ lattices is about $0.69$ for the 1D chain and shows no significant change with the increase of lattice width. The difference of the decaying behaviour between this two types of nonlinear lattices may result from the harmonic interactions in the $\text{FPU-}\beta$ model, where phonon has been excited and may contribute to heat diffusion.

\begin{figure}[ht]
\includegraphics[width=0.8 \columnwidth]{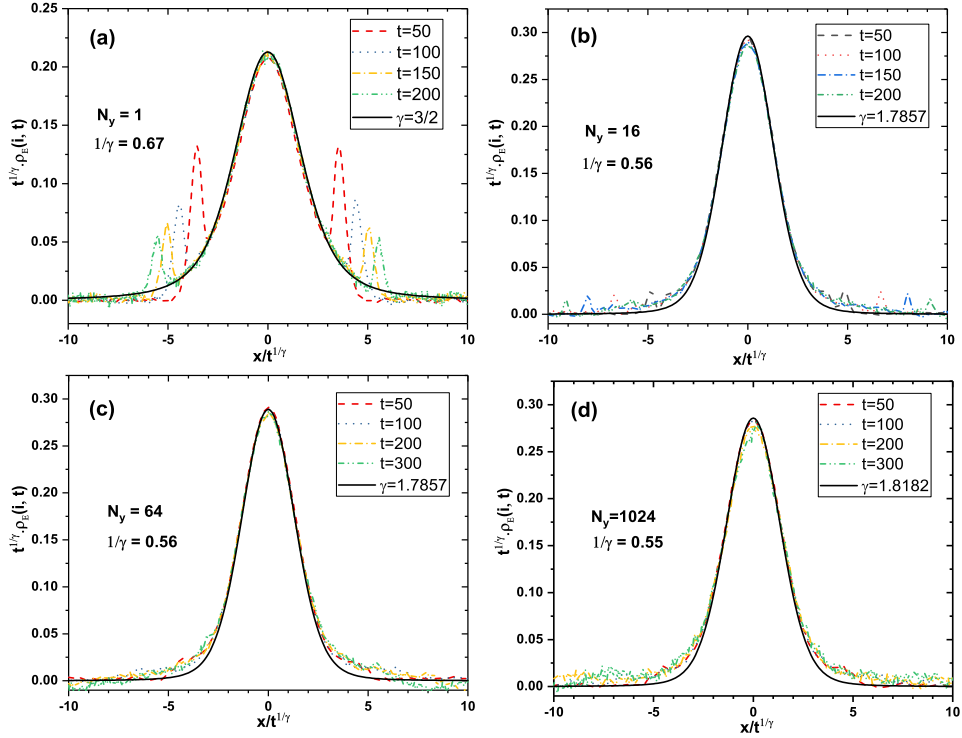}
\caption{\label{fig5:Equartic_rescale} (Color online). The rescaled energy fluctuation correlation function $t^{1/\gamma}\rho_{E}(i/t^{1/\gamma}, t)$ in the purely quartic lattices with the width (a) $N_y=1$, (b) $N_y=16$, (c) $N_y=64$, (d) $N_y=1024$. Here the exponent $\gamma$ is obtained by fitting the decay height of the energy fluctuation correlation function to the power-law distribution $\sim t^{-1/\gamma}$ shown in Fig.~\ref{fig4:Eh_decay}.   The black solid line in the figure denotes the fit to the L\'{e}vy-stable distribution $f_{LW}^{\gamma}(x)$.  }
\end{figure}

\begin{figure}[ht]
\includegraphics[width=0.8 \columnwidth]{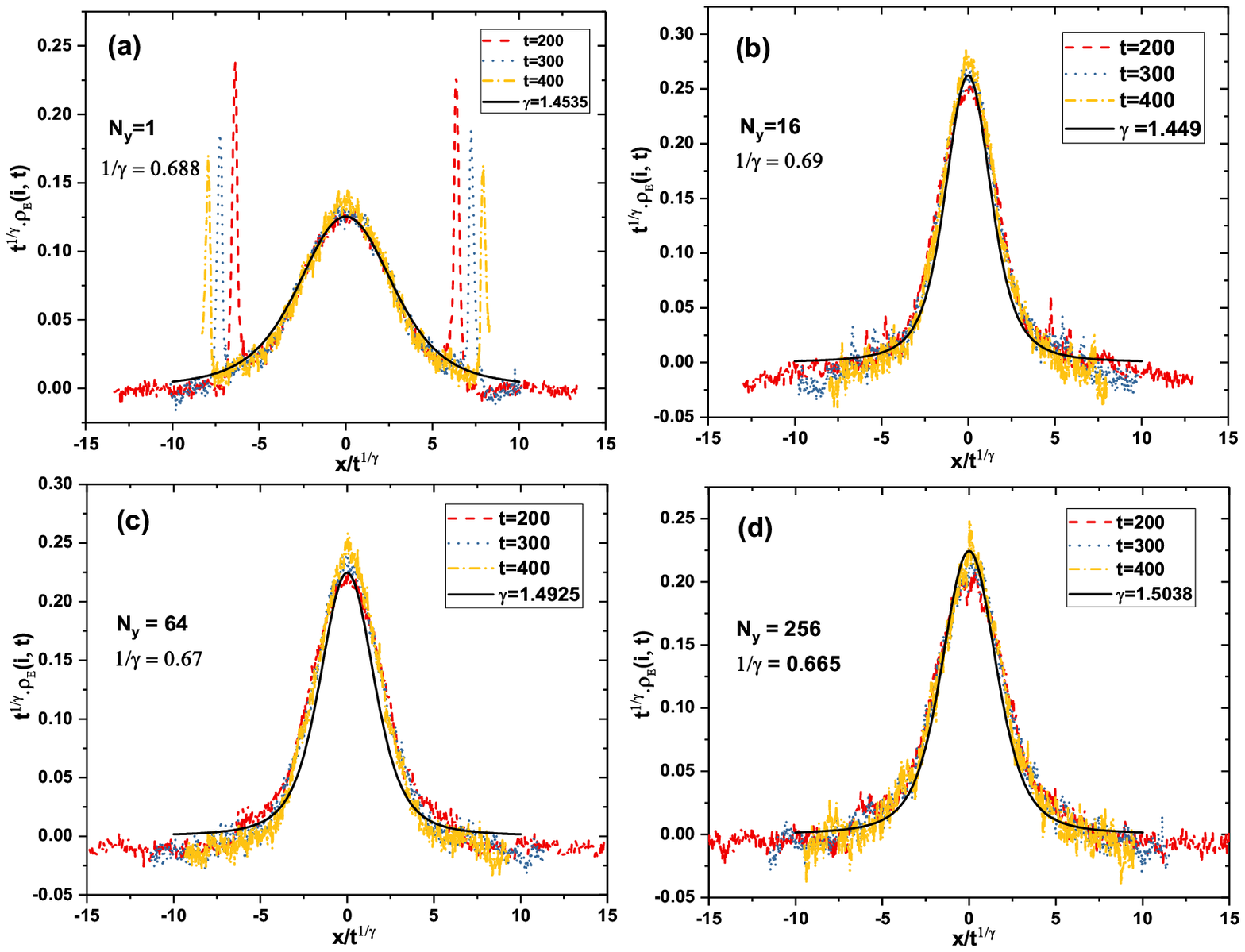}
\caption{\label{fig6:Efpu_rescale} (Color online). The rescaled energy fluctuation correlation function $t^{1/\gamma}\rho_{E}(i/t^{1/\gamma}, t)$ in  the $\text{FPU-}\beta$ lattices  with the width (a) $N_y=1$, (b) $N_y=16$, (c) $N_y=64$, (d) $N_y=1024$. Here the exponent $\gamma$ is obtained by fitting the decay height of the energy fluctuation correlation function to the power-law distribution $\sim t^{-1/\gamma}$ shown in Fig.~\ref{fig4:Eh_decay}.   The black solid line in the figure denotes the fit to the L\'{e}vy-stable distribution $f_{LW}^{\gamma}(x)$.  }
\end{figure}

 Furthermore, to identify the diffusion property of heat modes, the rescaled energy fluctuation correlation functions $t^{1/\gamma}\rho_{E}(i,t)\sim i/t^{1/\gamma}$ with the varying width for the purely quartic lattices and the $\text{FPU-}\beta$ model are depicted in Fig.~\ref{fig5:Equartic_rescale} and Fig.~\ref{fig6:Efpu_rescale}, respectively. The scaling exponent $\gamma$ is obtained from the relation $H_c^E \sim t^{-1/\gamma}$ shown in Fig.~\ref{fig4:Eh_decay}. To further confirm the superdiffusive property of $\rho_{E}(i,t)$, the L\'{e}vy-stable distribution\cite{Zaburdaev2015,firststeprw} $f_{LW}^{\gamma}(i,t)$ with the  same scaling exponent $\gamma$ is also plotted by the solid line in each figure. From the the figure, we can find that the collapse of the central heat mode at different time after scaling is very good according to the scaling invariant in Eq.~(\ref{scaling}), irrespective of the lattice width $N_y$. The fitting to the corresponding L\'{e}vy-stable distribution is also quite fine.  Besides, as can be seen from Fig.~\ref{fig5:Equartic_rescale}(a) and Fig.~\ref{fig6:Efpu_rescale}(b), the side peak in the 1D system does not follow the same scaling property of heat mode as predicted by the nonlinear fluctuating hydrodynamic theory\cite{Spohn2014,Das2014a}. Here, we focus on the scaling property of the heat modes. The nonlinear hydrodynamic fluctuation theory\cite{Beijeren2012,Spohn2014,Das2014a} predicts that the scaling exponent $\gamma$ scales according to the L\'{e}vy-3/2 distribution (i.e. $\gamma=3/2$) for the 1D momentum-conserving nonlinear lattices with an even potential at zero pressure, like the purely quartic lattice and the $\text{FPU-}\beta$ model. As indicated in the figure, our obtained values of the scaling exponent $\gamma$ agree well with the prediction of nonlinear hydrodynamic fluctuation theory, with $\gamma=3/2$ for the 1D purely quartic lattice and $\gamma=1.454$ for  the $\text{FPU-}\beta$ model. When the lattice width $N_y$ is increased to $1024$, the scaling exponent $\gamma$ reaches $\gamma=1.818$ for the 1D purely quartic lattice and $\gamma=1.504$ for the $\text{FPU-}\beta$ model, respectively. All values of the scaling exponent $\gamma$ with different width fall into the range $1<\gamma<2$, thus confirming the superdiffusive property of heat modes for both the purely quartic 2D lattice and the $\text{FPU-}\beta$ 2D model. Our simulations exhibit a slower energy diffusion with the dimensional crossover from 1D to 2D in the momentum-conserving nonlinear lattices.

 Finally, as can be seen in Fig.~\ref{fig3:Efei4_diff}, a Gaussian distribution function(normal diffusion) of heat mode can be observed for the $\phi^4$ model with an on-site potential, which breaks the conservation of momentum. Its profile of energy fluctuation correlation function $\rho_{E}$ can be perfectly well described by the Gaussian distribution $\rho_{E} \sim e^{-i^2/{4\pi D_Et}}/\sqrt{4\pi D_Et}$, where $D_E$ denotes the diffusion constant for the heat mode. The normal heat diffusion has been well verified\cite{H.Zhao2006,phi4book} in the 1D lattices with a $\phi^4$ potential. After numerically examining the Gaussian property of heat mode in Fig.~\ref{fig3:Efei4_diff}, we find that the energy fluctuation correlation function $\rho_{E}$ satisfies the Gaussian distribution both for 1D and 2D $\phi^4$ model, independent of the lattice width. Actually, this result is obvious and generally assumed beyond any doubt\cite{H.Zhao2006,phi4book}.

 To sum up, we have investigated the spatiotemporal profiles of energy fluctuation correlation, which can manifest the superdiffusion of energy in the 2D momentum-conserving nonlinear lattices. In addition to this, the mean square deviation(MSD)of energy distribution $\langle\Delta x^{2}(i,t)\rangle_{E}$ (i.e. the second moments of energy in terms of the probability function) can explicitly characterize energy diffusion in the asymptotic time limit and is related to heat conduction. Next, we  we quantitatively characterize heat transport by calculating the MSD of energy distribution$\langle\Delta x^{2}(t)\rangle_{E}$ given by  Eq.(\ref{correlation}).

\subsection{Mean square deviation(MSD)of energy distribution}
\begin{figure}[htb]
\includegraphics[width=0.8 \columnwidth]{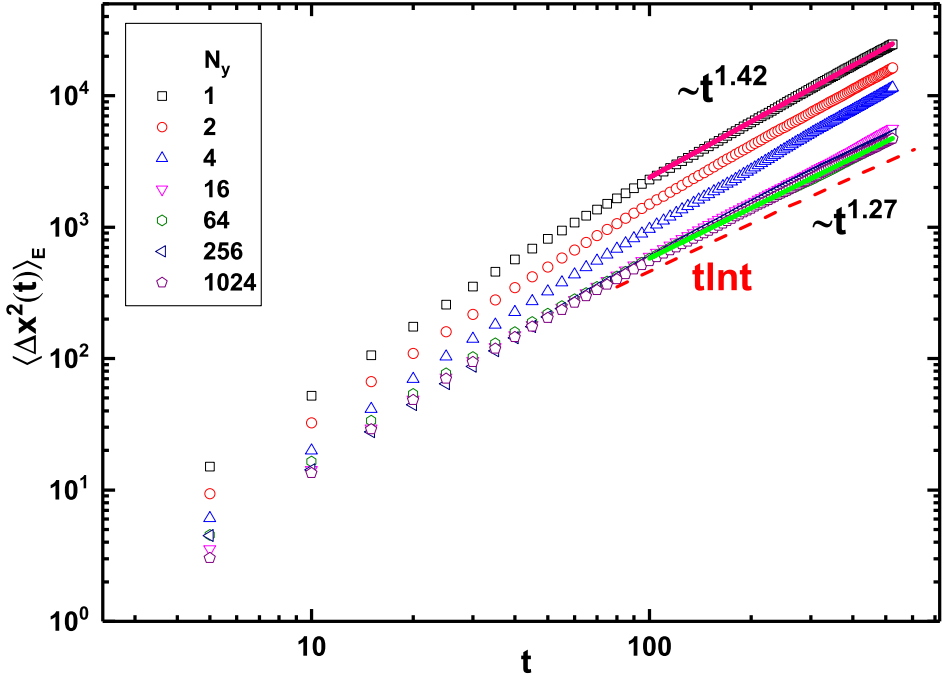}
\caption{\label{fig7:Equartic_coeff} (Color online).  The mean-square deviation(MSD) of energy $\langle\Delta x^{2}(t)\rangle_{E}$ as a function of time for the purely quartic lattices with width $N_y$  varying from 1 to 1024 as shown in the figure. The fittings of MSD to a power-law distribution $\sim t^{\beta}$  for the lattices of the width $N_y=1$ and $N_y=1024$ are plotted in the figure, respectively. The red dash line in the figure represents the function of $t\ln t$.    }
\end{figure}
\begin{figure}[htb]
\includegraphics[width=0.8 \columnwidth]{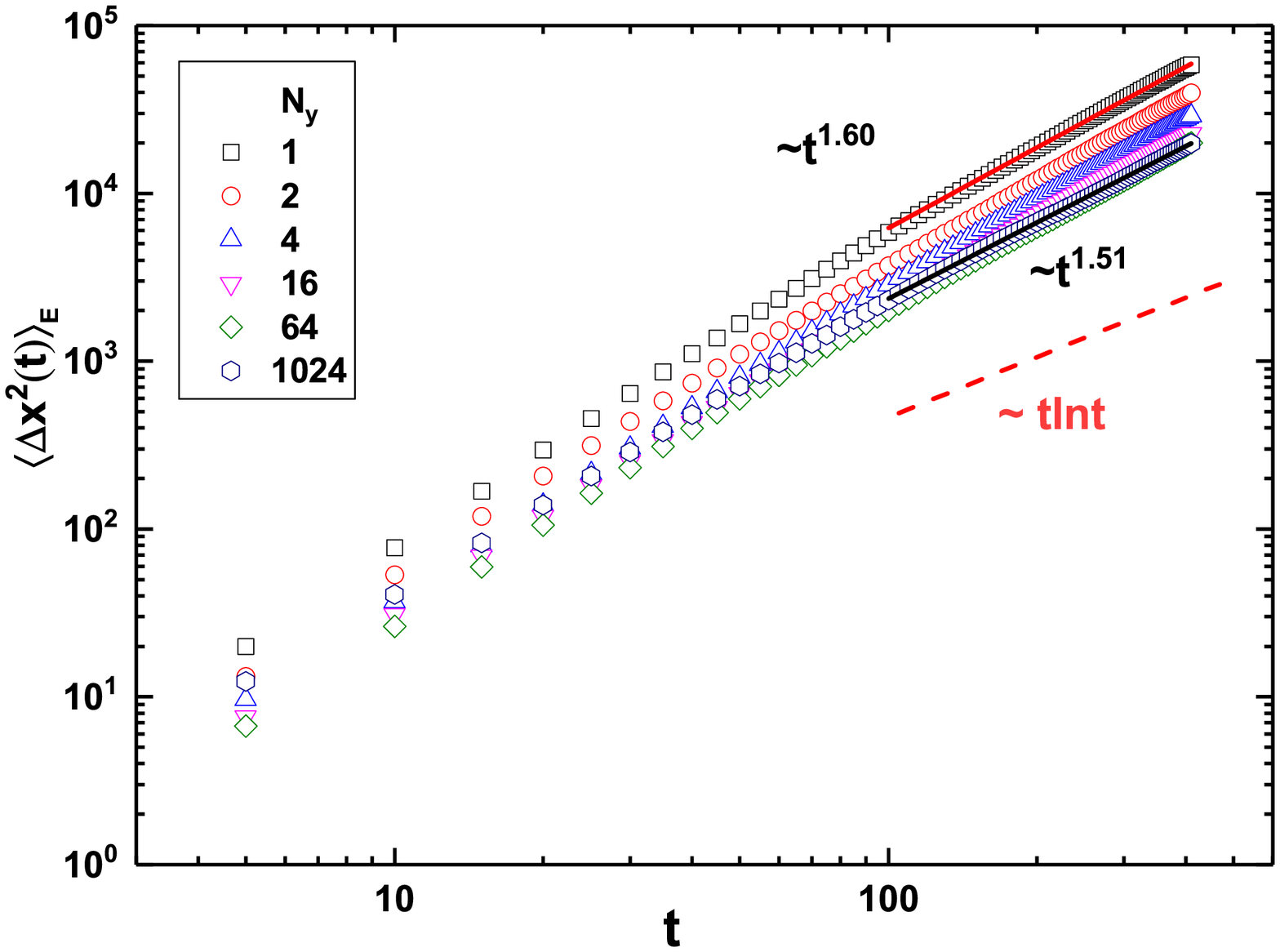}
\caption{\label{fig8:Efpu_coeff} (Color online).  The mean-square deviation(MSD) of energy $\langle\Delta x^{2}(t)\rangle_{E}$ as a function of time for the $\text{FPU-}\beta$ lattices with width $N_y$  varying from 1 to 1024 as shown in the figure. The fittings of MSD to a power-law distribution $\sim t^{\beta}$  for the lattices of the width $N_y=1$ and $N_y=1024$ are plotted in the figure, respectively. The red dash line in the figure represents the function of $t\ln t$.     }
\end{figure}
\begin{figure}[htb]
\includegraphics[width=0.8 \columnwidth]{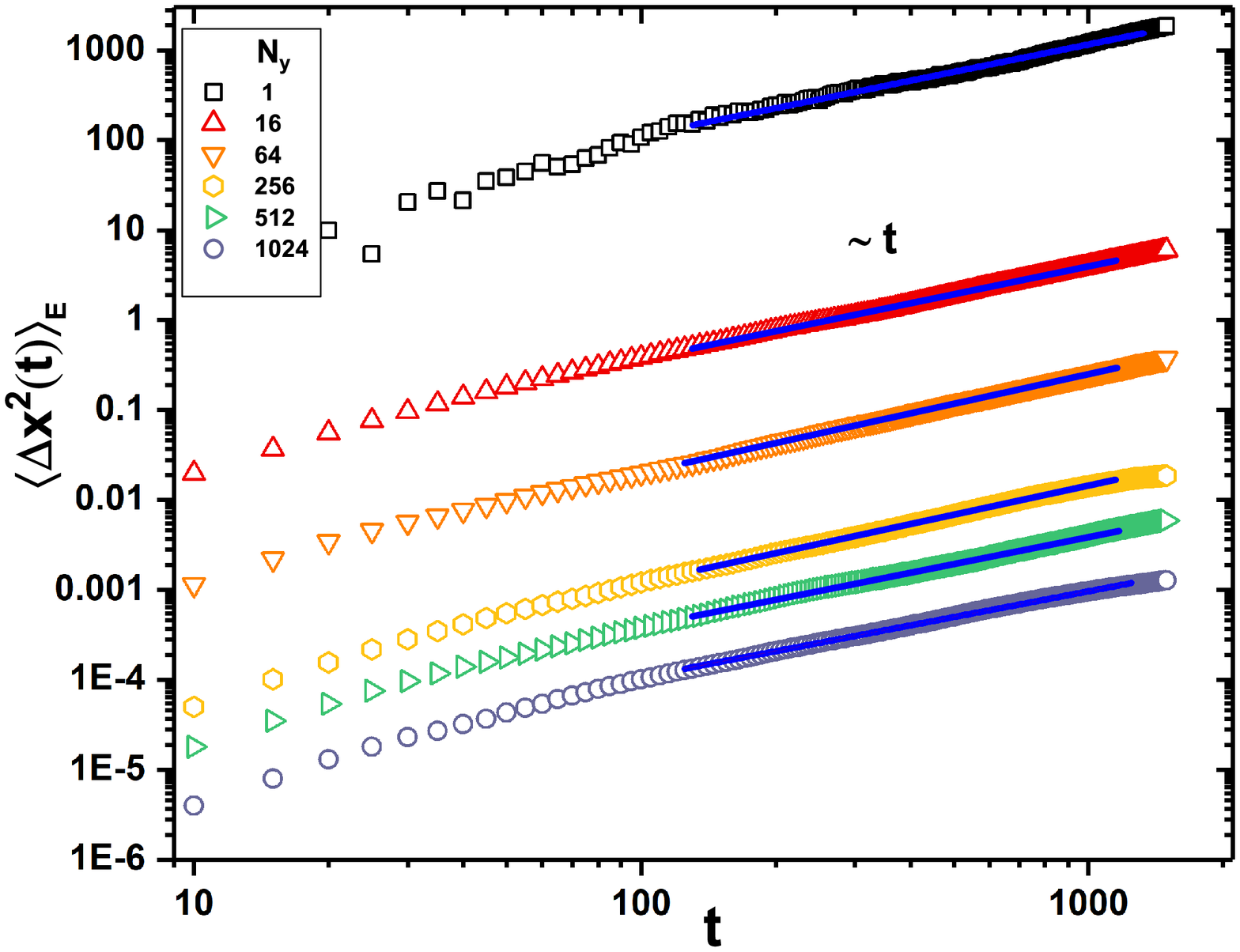}
\caption{\label{fig9:Ephi4_coeff} (Color online).  The mean-square deviation(MSD) of energy $\langle\Delta x^{2}(t)\rangle_{E}$ as a function of time for the  $\phi^4$  lattices with width $N_y$  varying from 1 to 1024 as shown in the figure. The fittings of MSD to a proportional relationship $\sim t$  for the lattices of the width $N_y=1$ and $N_y=1024$ are plotted in the figure, respectively.     }
\end{figure}
Heat conduction in 1D system can be directly related to energy diffusion process. According to the connection theory\cite{Liu2014b}, the MSD in 1D system obeys the second order differential equation with time as
\begin{equation}
\label{msdconnect}
 \frac{d^2\langle\Delta x^{2}(t)\rangle_{E}}{dt^2} =\frac{2C_{JJ}(t)}{k_BT^2c},
\end{equation}
 where $c$ is the specific heat capacity and $k_B$ is the Boltzmann constant. The autocorrelation function of heat currents $C_{JJ}(t)$ is related to thermal conductivity $\kappa$ through the Green-Kubo formula\cite{mathp2000,Lepri2003c,Dhar2008} given by $\kappa=\frac{1}{k_BT^2}\int^{\infty}_{0}dtC_{JJ}(t)$.  If the mean square deviation(MSD) of a conserving energy distribution $\langle\Delta x^{2}(t)\rangle_{E}$ is assumed to scale as $\langle\Delta x^{2}(t)\rangle_{E} \sim  t^{\beta}$, the behavior of diffusion process for a conservation quantity can be categorized\cite{Zaburdaev2015,Liu2014b} with regard to the exponent $\beta$: the normal diffusion when $\beta=1$, the superdiffusion when $\beta>1$, and the subdiffusion when $\beta<1$. For example, the MSD $\langle\Delta x^{2}(t)\rangle_{E}$ will  be proportional to time $t$, i.e. $\langle\Delta x^{2}(t)\rangle_{E} \sim  t$ when the energy fluctuation correlation function $\rho_{E}$ is the Gaussian distribution like the $\phi^4$ model. Thus, the normal energy diffusion will give rise to finite thermal conductivity, indicating normal heat conduction. Furthermore, if the profile of energy fluctuation correlation function $\rho_{E}$ belongs to the L\'{e}vy-stable distribution, the MSD will spread faster than the normal diffusion and exhibits the superdiffusion with $\beta=3-\gamma$ given by the L\'{e}vy walk theory\cite{Zaburdaev2015,firststeprw}. Anomalous heat conduction has been verified in the 1D momentum-conserving nonlinear lattices\cite{H.Zhao2006,Chen2013,Li2015,Gao2016c,xiong2017,Xiong2018a,Xiong2018b} such as the $\text{FPU-}\beta$ and purely quartic lattices. This relation\cite{Liu2014b} of energy diffusion to heat transport has been quantitatively verified in 1D nonlinear with symmetrical potential. As for 2D nonlinear lattices, there is no strict theoretical relationship between energy diffusion and heat transport, but we can still employ energy diffusion to qualitatively identify whether heat transport is normal or anomalous.

 The calculated distributions of MSD energy $\langle\Delta x^{2}(t)\rangle_{E}$  versus evolving time $t$ are plotted in Fig.~\ref{fig7:Equartic_coeff}, Fig.~\ref{fig8:Efpu_coeff} and Fig.~\ref{fig9:Ephi4_coeff} for the purely quartic lattices, the $\text{FPU-}\beta$ lattices and the $\phi^4$ model with the varying width, respectively.  As can be seen from Fig.~\ref{fig7:Equartic_coeff} and Fig.~\ref{fig8:Efpu_coeff}, the MSD of energy on the purely quartic and $\text{FPU-}\beta$ model shows the superdiffusion behavior. We can find that the diffusion exponent $\beta$ decreases with the increase of lattice width. For example, the diffusion exponent $\beta$ for the purely quartic lattices is reduced from $\beta=1.42$ for $N_y=1$ to $\beta=1.27$ for $N_y=1024$. The obtained $\beta$ of $1.42$ for the 1D chain $N_y=1$ is consistent with the previously reported\cite{H.Zhao2006,Chen2013,Li2015} diffusion coefficient $1.4$. When the width is increased to $1024$, the diffusion exponent decreases to  $\beta$ of $1.27$ as shown in Fig.~\ref{fig7:Equartic_coeff}, still implying an anomalous heat conduction in the 2D quartic lattices.  As for the 2D systems, the mode-coupling theory\cite{Lepri2003c,Dhar2008} predicts that the autocorrelation function of heat currents $C_{JJ}(t)$ decays as $t^{-1}$, which leads to the logarithmic divergence of thermal conductivity with system size $N$ as $\kappa\propto \ln N$. If the connection theory Eq.~(\ref{msdconnect}) is still  valid in 2D systems, the MSD of energy will change over time as $\langle\Delta x^{2}(t)\rangle_{E} \sim  t\ln t$. To compare with the power-law relationship, we have plotted the function of $t\ln t$ in Fig.~\ref{fig7:Equartic_coeff} using the red dash line. As illustrated in the figure, the discrepancy in the scaling behavior between the power-law $\sim t^{1.27}$ and the function of $\sim t\ln t$ is so small that it is difficult to numerically determine whether the observed MSD of energy distribution conforms to the power-law $\sim t^{1.27}$ or to the function of $\sim t\ln t$ in the asymptotic time limit. A logarithmically divergent thermal conductivity with the system size $N$ is recently reported\cite{Wang2012a} in the purely quartic lattices with a scalar displacement field. Note that our purely quartic lattice has a vector displacement. Considering the sound numerical fitting as the power-law in Fig.~\ref{fig7:Equartic_coeff}, our results of energy diffusion may be in favor of a logarithmic divergence of thermal conductivity with size in the purely quartic lattices with a vector displacement. At the same time, we cannot exclude that this may be a numerical coincidence. But, the calculated distributions of MSD energy $\langle\Delta x^{2}(t)\rangle_{E}$ for energy diffusion in Fig.~\ref{fig7:Equartic_coeff} assuredly confirms that heat transport in the 2D purely quartic lattices is anomalous. In addition, it can be seen from the figure that the dimensional crossover from 1D to 2D with the varying width occurs rapidly such that the diffusion exponent $\beta$ for $N_y=16$ almost converges to the value of $1.27$. This rapidly convergent tendency is consistent with the direct simulation of thermal conductivity\cite{Savin2016}, where the converged results are obtained when the width  $N_y > 32$.

 Similar superdiffusive time dependence of the MSD energy distribution can be observed for the $\text{FPU-}\beta$ model in Fig.~\ref{fig8:Efpu_coeff}. In comparison with the purely quartic lattices, the $\text{FPU-}\beta$ model introduces harmonic interaction, which has produced a stronger coupling between the heat and sound modes in the 1D nonlinear chain as indicated by the side peaks in Fig.~\ref{fig2:Ebeta_diff}(a). This coupling of the energy and momentum diffusion may contribute to energy diffusion in the $\text{FPU-}\beta$ lattices. A fast superdiffusive energy spreading, growing as $\langle\Delta x^{2}(t)\rangle_{E} \sim  t^{1.60}$, is observed for the 1D lattices in Fig.~\ref{fig8:Efpu_coeff}. When the width is increased to $1024$, the diffusion exponent $\beta$ converges rapidly to $\beta=1.51$. To compare with the prediction of logarithmical divergency by the mode-coupling theory, a function of $\sim t\ln t$ is also illustrated in Fig.~\ref{fig8:Efpu_coeff} by the red dash line. By contrast, a significant difference exits between the power-law scale $\sim \beta=1.51$ and the function of $\sim t\ln t$. Therefore, a power-law relationship can be assumed for the anomalous energy diffusion in the 2D $\text{FPU-}\beta$ lattices. If the connection theory of Eq.~(\ref{msdconnect}) is still valid in the 2D system, our energy diffusion yields that thermal conductivity $\kappa$ will diverge as $\kappa \sim N^{0.51}$, which qualitatively coincides with the power-law divergence of thermal conductivity\cite{Shiba2008a} of the 2D $\text{FPU-}\beta$ lattices with a vector displacement by the direct simulation of thermal conductivity.
 In contrast, as shown in Fig.~\ref{fig9:Ephi4_coeff}, a normal MSD energy diffusion $\langle\Delta x^{2}(t)\rangle_{E} \sim  t$ can be observed in the $\phi^4$ lattices with a on-site potential, irrespective of the varying width. Put differently, we extends the conclusion from 1D to 2D systems that nonlinear lattices with a $\phi^4$ on-site potential will exhibit normal energy diffusion and thus give rise to Fourier law of heat transport.  The normal energy diffusion in the $\phi^4$ lattices obviously originates from the explicit Gaussian distributions of energy fluctuation correlation functions,as depicted in Fig.~\ref{fig3:Efei4_diff}. To sum up,  our MSD energy distribution strongly confirms anomalous energy diffusion in both the 1D and 2D momentum-conserving lattices of the purely quartic and $\text{FPU-}\beta$ model and normal energy diffusion in the $\phi^4$ 2D system.  Next, we turn to the characteristics of momentum diffusion in the 2D lattices.

\subsection{ Momentum diffusion }

\begin{figure}[ht]
\includegraphics[width=0.8 \columnwidth]{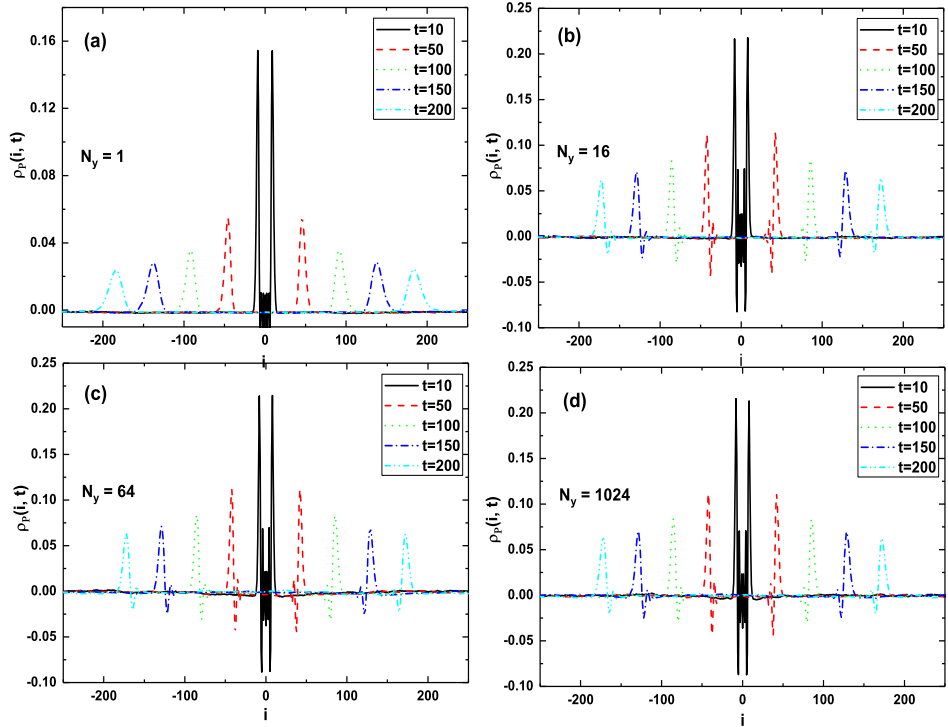}
\caption{\label{fig10:momentum_quartic_diff} (Color online).  The spatial profiles of the momentum fluctuation correlation function $\rho_{P}(i,t)$ for the purely quartic lattices with the increasing width (a) $N_y=1$, i.e. 1D lattices, (b) $N_y=16$, (c) $N_y=64$, (d) $N_y=1024$. The spatial profiles of $\rho_{P}(i,t)$ at times $ t=10,50,100,150, \text{and} \; 200 $ are labelled by different lines shown in the figure.   }
\end{figure}

\begin{figure}[ht]
\includegraphics[width=0.8 \columnwidth]{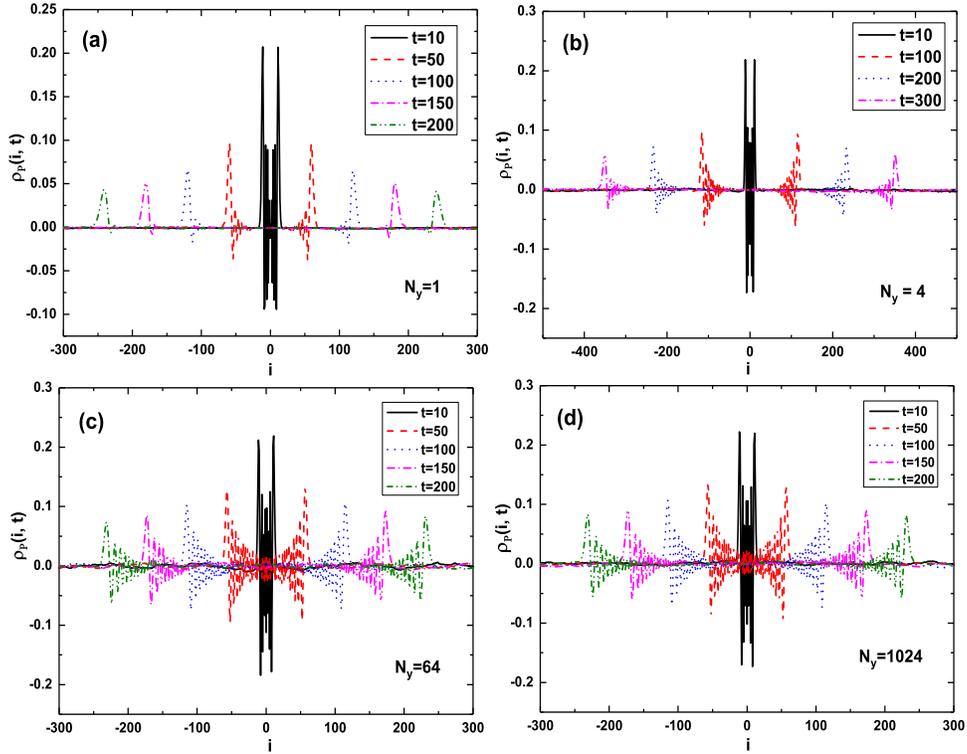}
\caption{\label{fig11:momentum_fpubeta_diff} (Color online). The spatial profiles of the momentum fluctuation correlation function $\rho_{P}(i,t)$ for the $\text{FPU-}\beta$ lattices with the increasing width (a) $N_y=1$, i.e. 1D lattices, (b) $N_y=16$, (c) $N_y=64$, (d) $N_y=1024$. The spatial profiles of $\rho_{P}(i,t)$ at times $ t=10,50,100,150, \text{and} \; 200 $ are labelled by different lines shown in the figure.  }
\end{figure}

According to the nonlinear fluctuating hydrodynamic theory\cite{Beijeren2012,Spohn2014,Das2014a}, the full three normal modes (one heat mode and two sound modes) have determined heat transport in 1D nonlinear lattice. Sound mode or momentum diffusion can be related\cite{Beijeren2012,Spohn2014,Das2014a} to the momentum fluctuation correlation.  Among the three types of nonlinear lattices considered,  momentum diffusion cannot be well defined in the $\phi^4$ model, owing to its non-conserving  momentum. Therefore, we focus on the behavior of momentum diffusion in the purely quartic lattices and the $\text{FPU-}\beta$ lattices. Momentum diffusion is represented by the momentum fluctuation correlation function $\rho_{P}(i,t)$ defined in Eq.~(\ref{correlation}). The spatial profiles of momentum fluctuation correlation function $\rho_{P}$ for different times with varying lattice width for the purely quartic lattices and the $\text{FPU-}\beta$ lattices are depicted in Fig.~\ref{fig10:momentum_quartic_diff} and Fig.~\ref{fig11:momentum_fpubeta_diff}, respectively. It is interesting to investigate the dimensional-crossover features of momentum diffusion from 1D to 2D system. It can be seen from the figure that two side peaks of sound mode  move ballistically outward with a constant sound velocity $c$. We have numerically fitted the sound speed from the profiles of $\rho_{P}(i,t)$ at different correlation times, obtaining $c=0.914, 0.861$ for the purely quartic lattices with $N_y=1, 1024$ and $c=1.22, 1.16$ for the $\text{FPU-}\beta$  lattices with $N_y=1, 1024$, respectively. A small decrease of sound speed can be observed with the increase of lattice width. This decrease of sound speed may originate from the reduction of phonon group velocity because the phonon dispersion relations will change with the increase of lattice width.  Besides, the sound speed of the $\text{FPU-}\beta$ lattices is a bit larger than that of the purely quartic lattices, as a result of the harmonic interaction in the $\text{FPU-}\beta$ lattices.

In particular, as the lattice width increases, we can observe a significant oscillation in the diffusion of sound modes, especially for the $\text{FPU-}\beta$ lattices in Fig.~\ref{fig11:momentum_fpubeta_diff}. Similar oscillations have been displayed\cite{Xiong2018b} in the nonlinear chain with a harmonic on-site potential. The oscillations of ballistic spread of sound modes in 2D nonlinear lattice may indicate a competition between harmonic phonons and nonlinear interactions. It is well known that the harmonic interactions produce harmonic phonons, which are scattered by nonlinear interactions. However, such nonlinear phonon-phonon scattering is limited by the selection rules like the energy and momentum conservation. In 1D system, the selection rule for the conservation of both energy and momentum is difficult to be satisfied\cite{mathp2000,Lepri2003c,Dhar2008} at the same time. Therefore, the sound mode propagates without obvious decay as shown in Fig.~\ref{fig11:momentum_fpubeta_diff}(a). By contrast, these selection rules can be easily satisfied in 2D nonlinear lattices owing to the enlarged phase space, when the lattice width grows. Thus, in comparison with 1D nonlinear lattices, the sound modes in 2D nonlinear lattices will decay faster  as shown in Fig.~\ref{fig11:momentum_fpubeta_diff}(b-d).

\begin{figure}[ht]
\includegraphics[width=0.8 \columnwidth]{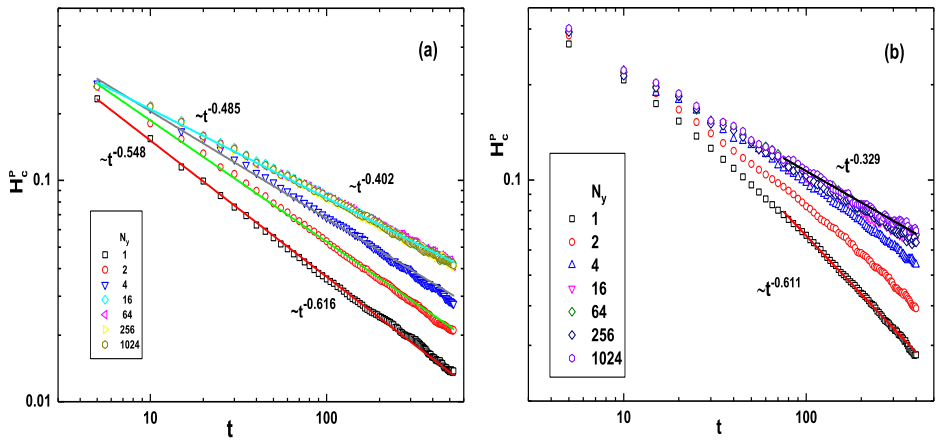}
\caption{\label{fig12:momentum_coeff} (Color online). The decay of the peak height $H^{P}_{c}$ of the momentum fluctuation correlation function as a function of time (a) in the 2d purely quartic lattice, (b) in the $\text{FPU-}\beta$ lattices. The decay of height is fitted to a power-law distribution $H^{P}_{c}\sim t^{-\mu}$ as shown in the figure.  }
\end{figure}

To further investigate the momentum diffusion, we have also calculated the decay of the peak height of the momentum fluctuation correlation function. The decay of the peak height of the momentum fluctuation correlation function as a function of time is depicted in Fig.~\ref{fig12:momentum_coeff} for the 2d purely quartic lattice and  the $\text{FPU-}\beta$ lattice, respectively. It can be seen from the figure that the decay of momentum height can be fitted to a power-law distribution $H^{P}_{c}\sim t^{-\mu}$, which is similar to that of energy.  A normal diffusion of momentum is characterized by the value of the decay exponent $\mu=0.5$. For the 1D system, both the purely quartic lattice and the $\text{FPU-}\beta$ lattice demonstrate a superdiffusion of momentum, with a decay exponent $\mu=0.616, 0.611$ correspondingly. Here, our result for the 1D nonlinear lattices is consistent with the previous hypothesis\cite{Li2015} that anomalous heat transport in  1D momentum-conserving Hamiltonian lattice systems is corroborated by the superdiffusive spreads of momentum excess density. However, when the lattice width increases, the momentum diffusion induces a crossover from superdiffusion in 1D system to normal or subdiffusion in 2D systems. For example, in case of the purely quartic lattice, the decay exponent $\mu$ changes from $0.616$ for the 1D system ($N_y=1$) to  $0.402$ for the 2D system ($N_y=1024$). Similar behaviors can also be observed in the $\text{FPU-}\beta$ lattices.

\section{Conclusions and discussion}
In summary, we have successfully investigated energy/momentum diffusions from 1D to 2D in the three types of nonlinear lattices: the purely quartic lattice, the $\text{FPU-}\beta$ lattice and the momentum-nonconserving $\phi^4$ model. The apparent advantage of the approach of the fluctuation correlation function is that it circumvents the finite-size problem during the direct simulation of heat conduction. We demonstrate that energy diffusion of the momentum-conserving 2D nonlinear lattices is anomalous and can be well fitted by the L\'{e}vy-stable distribution.  In contrast to the 1D energy fluctuation correlation function, the disappearance of the side peak in the momentum-conserving 2D nonlinear lattices indicates a weak coupling between heat mode and sound mode, thus implying a reduced heat transport. Another approach of the mean-square deviation of energy further confirms the anomalous diffusion of energy in the momentum-conserving 2D nonlinear lattices. Compared with the purely quartic lattice, a faster energy diffusion is observed in the $\text{FPU-}\beta$ lattice, owing to the harmonic interactions. Besides, the simulations of momentum diffusion illustrate that two sound modes in the 2D nonlinear move ballistically outward with a constant sound velocity, similar to that of the 1D chain. However, with the increase of lattice width, we can observe a significant oscillation of diffusion in the 2D sound modes compared with that of the 1D system, possibly as a result of competition of harmonic phonons and nonlinear interactions. Contrary to the hypothesis of the 1D system, we find that anomalous heat transport in the 2D momentum-conserving Hamiltonian lattice systems is not accompanied by the superdiffusion of momentum any more. Moreover, as is expected, we find that nonlinear lattices with a $\phi^4$ on-site potential exhibit normal energy diffusion, independent of its dimension.

Our studies here confirm that energy diffusion in the momentum-conserving 2D nonlinear lattices is anomalous. The anomalous energy diffusion can ensure anomalous heat conduction in the 2D nonlinear lattices. How to relate energy diffusion to heat conduction will be an important and interesting problem. A connection relationship for the 1D chain system has been proposed in Eq.~(\ref{msdconnect}).  Nevertheless, a theoretical relationship between energy diffusion and heat conduction in 2D system is still lacking so far.  If we assume that the connection relation for the 1D system is still valid for the 2D system, the scaling of MSD of energy $ \sim t\ln t$ for the 2d purely quartic lattice may suggest the logarithmic divergence of thermal conductivity with system size and the power-law scaling for the 2D $\text{FPU-}\beta$ lattice implies a power-law divergence of thermal conductivity. Based on the simulations, we infer that the scaling behavior of divergence of thermal conductivity in the 2D nonlinear lattices may depend on the competition of harmonic and anharmonic interactions: the purely anharmonic interactions will lead to a logarithmic divergence of thermal conductivity with system size while the harmonic interaction will contribute to a power-divergence of thermal conductivity. However, further investigations both theoretical and numerical are needed. Our present simulations expound the scaling behaviour of energy/momentum diffusion in the 2D nonlinear lattices. We hope our results contribute to understanding heat transport in 2D nonlinear lattices.

\section{Acknowledgement}
We thank Daxing Xiong, Baowen Li, and Yong Zhang for useful discussions. J. Wang acknowledges the support from National Natural Science
Foundation of China (NSFC) under the grant 11875047 and 11375148. N.B. Li acknowledges the support from National Natural Science
Foundation of China (NSFC) under the grant 11775158.

\newpage

\bibliography{t2}

\begin{thebibliography}{46}
\expandafter\ifx\csname natexlab\endcsname\relax\def\natexlab#1{#1}\fi
\expandafter\ifx\csname bibnamefont\endcsname\relax
  \def\bibnamefont#1{#1}\fi
\expandafter\ifx\csname bibfnamefont\endcsname\relax
  \def\bibfnamefont#1{#1}\fi
\expandafter\ifx\csname citenamefont\endcsname\relax
  \def\citenamefont#1{#1}\fi
\expandafter\ifx\csname url\endcsname\relax
  \def\url#1{\texttt{#1}}\fi
\expandafter\ifx\csname urlprefix\endcsname\relax\def\urlprefix{URL }\fi
\providecommand{\bibinfo}[2]{#2}
\providecommand{\eprint}[2][]{\url{#2}}

\bibitem[{\citenamefont{Bonetto et~al.}(2000)\citenamefont{Bonetto, Lebowitz,
  and Rey-Bellet}}]{mathp2000}
\bibinfo{author}{\bibfnamefont{F.}~\bibnamefont{Bonetto}},
  \bibinfo{author}{\bibfnamefont{J.}~\bibnamefont{Lebowitz}}, \bibnamefont{and}
  \bibinfo{author}{\bibfnamefont{L.}~\bibnamefont{Rey-Bellet}},
  \emph{\bibinfo{title}{Mathematical Physics 2000, p. 128}}
  (\bibinfo{publisher}{Imperial College Press, London}, \bibinfo{year}{2000}).

\bibitem[{\citenamefont{Lepri et~al.}(2003)\citenamefont{Lepri, Livi, and
  Politi}}]{Lepri2003c}
\bibinfo{author}{\bibfnamefont{S.}~\bibnamefont{Lepri}},
  \bibinfo{author}{\bibfnamefont{R.}~\bibnamefont{Livi}}, \bibnamefont{and}
  \bibinfo{author}{\bibfnamefont{A.}~\bibnamefont{Politi}},
  \bibinfo{journal}{Phys. Rep.} \textbf{\bibinfo{volume}{377}},
  \bibinfo{pages}{1} (\bibinfo{year}{2003}).

\bibitem[{\citenamefont{Dhar}(2008)}]{Dhar2008}
\bibinfo{author}{\bibfnamefont{A.}~\bibnamefont{Dhar}}, \bibinfo{journal}{Adv.
  Phys.} \textbf{\bibinfo{volume}{57}}, \bibinfo{pages}{457}
  (\bibinfo{year}{2008}).

\bibitem[{\citenamefont{Wang et~al.}(2008)\citenamefont{Wang, Wang, and
  Lü}}]{Wang2008a}
\bibinfo{author}{\bibfnamefont{J.-S.} \bibnamefont{Wang}},
  \bibinfo{author}{\bibfnamefont{J.}~\bibnamefont{Wang}}, \bibnamefont{and}
  \bibinfo{author}{\bibfnamefont{J.~T.} \bibnamefont{Lü}},
  \bibinfo{journal}{Eur. Phys. J. B} \textbf{\bibinfo{volume}{62}},
  \bibinfo{pages}{381} (\bibinfo{year}{2008}).

\bibitem[{\citenamefont{Lepri}(2016)}]{review2016}
\bibinfo{editor}{\bibfnamefont{S.}~\bibnamefont{Lepri}}, ed.,
  \emph{\bibinfo{title}{Thermal Transport in Low Dimensions}}, vol.
  \bibinfo{volume}{921} of \emph{\bibinfo{series}{Lecture notes in Physics}}
  (\bibinfo{publisher}{Springer, Cham}, \bibinfo{year}{2016}).

\bibitem[{\citenamefont{Li et~al.}(2012)\citenamefont{Li, Ren, Wang, Zhang,
  Hänggi, and Li}}]{Li2012a}
\bibinfo{author}{\bibfnamefont{N.}~\bibnamefont{Li}},
  \bibinfo{author}{\bibfnamefont{J.}~\bibnamefont{Ren}},
  \bibinfo{author}{\bibfnamefont{L.}~\bibnamefont{Wang}},
  \bibinfo{author}{\bibfnamefont{G.}~\bibnamefont{Zhang}},
  \bibinfo{author}{\bibfnamefont{P.}~\bibnamefont{Hänggi}}, \bibnamefont{and}
  \bibinfo{author}{\bibfnamefont{B.}~\bibnamefont{Li}}, \bibinfo{journal}{Rev.
  Mod. Phys.} \textbf{\bibinfo{volume}{84}}, \bibinfo{pages}{1045}
  (\bibinfo{year}{2012}).

\bibitem[{\citenamefont{Gu et~al.}(2018)\citenamefont{Gu, Wei, Yin, Li, and
  Yang}}]{Gu2018a}
\bibinfo{author}{\bibfnamefont{X.}~\bibnamefont{Gu}},
  \bibinfo{author}{\bibfnamefont{Y.}~\bibnamefont{Wei}},
  \bibinfo{author}{\bibfnamefont{X.}~\bibnamefont{Yin}},
  \bibinfo{author}{\bibfnamefont{B.}~\bibnamefont{Li}}, \bibnamefont{and}
  \bibinfo{author}{\bibfnamefont{R.}~\bibnamefont{Yang}},
  \bibinfo{journal}{Rev. Mod. Phys.} \textbf{\bibinfo{volume}{90}},
  \bibinfo{pages}{1103} (\bibinfo{year}{2018}).

\bibitem[{\citenamefont{Lepri et~al.}(1997)\citenamefont{Lepri, Livi, and
  Politi}}]{Lepri1997}
\bibinfo{author}{\bibfnamefont{S.}~\bibnamefont{Lepri}},
  \bibinfo{author}{\bibfnamefont{R.}~\bibnamefont{Livi}}, \bibnamefont{and}
  \bibinfo{author}{\bibfnamefont{A.}~\bibnamefont{Politi}},
  \bibinfo{journal}{Phys. Rev. Lett.} \textbf{\bibinfo{volume}{78}},
  \bibinfo{pages}{1896} (\bibinfo{year}{1997}).

\bibitem[{\citenamefont{Lepri et~al.}(1998)\citenamefont{Lepri, Livi, and
  Politi}}]{Lepri1998}
\bibinfo{author}{\bibfnamefont{S.}~\bibnamefont{Lepri}},
  \bibinfo{author}{\bibfnamefont{R.}~\bibnamefont{Livi}}, \bibnamefont{and}
  \bibinfo{author}{\bibfnamefont{A.}~\bibnamefont{Politi}},
  \bibinfo{journal}{Europhys. Lett.} \textbf{\bibinfo{volume}{43}},
  \bibinfo{pages}{271} (\bibinfo{year}{1998}).

\bibitem[{\citenamefont{Hu et~al.}(1998)\citenamefont{Hu, Li, and
  Zhao}}]{PhysRevE.57.2992}
\bibinfo{author}{\bibfnamefont{B.}~\bibnamefont{Hu}},
  \bibinfo{author}{\bibfnamefont{B.}~\bibnamefont{Li}}, \bibnamefont{and}
  \bibinfo{author}{\bibfnamefont{H.}~\bibnamefont{Zhao}},
  \bibinfo{journal}{Phys. Rev. E} \textbf{\bibinfo{volume}{57}},
  \bibinfo{pages}{2992} (\bibinfo{year}{1998}).

\bibitem[{\citenamefont{Lepri et~al.}(2005)\citenamefont{Lepri, Livi, and
  Politi}}]{Lepri2005}
\bibinfo{author}{\bibfnamefont{S.}~\bibnamefont{Lepri}},
  \bibinfo{author}{\bibfnamefont{R.}~\bibnamefont{Livi}}, \bibnamefont{and}
  \bibinfo{author}{\bibfnamefont{A.}~\bibnamefont{Politi}},
  \bibinfo{journal}{Chaos} \textbf{\bibinfo{volume}{15}},
  \bibinfo{pages}{015118} (\bibinfo{year}{2005}).

\bibitem[{\citenamefont{Mai et~al.}(2007)\citenamefont{Mai, Dhar, and
  Narayan}}]{Mai2007}
\bibinfo{author}{\bibfnamefont{T.}~\bibnamefont{Mai}},
  \bibinfo{author}{\bibfnamefont{A.}~\bibnamefont{Dhar}}, \bibnamefont{and}
  \bibinfo{author}{\bibfnamefont{O.}~\bibnamefont{Narayan}},
  \bibinfo{journal}{Phys. Rev. Lett.} \textbf{\bibinfo{volume}{98}},
  \bibinfo{pages}{184301} (\bibinfo{year}{2007}).

\bibitem[{\citenamefont{Hatano}(1999)}]{Hatano1999}
\bibinfo{author}{\bibfnamefont{T.}~\bibnamefont{Hatano}},
  \bibinfo{journal}{Phys. Rev. E} \textbf{\bibinfo{volume}{59}},
  \bibinfo{pages}{R1} (\bibinfo{year}{1999}).

\bibitem[{\citenamefont{Grassberger et~al.}(2002)\citenamefont{Grassberger,
  Nadler, and Yang}}]{Grassberger2002}
\bibinfo{author}{\bibfnamefont{P.}~\bibnamefont{Grassberger}},
  \bibinfo{author}{\bibfnamefont{W.}~\bibnamefont{Nadler}}, \bibnamefont{and}
  \bibinfo{author}{\bibfnamefont{L.}~\bibnamefont{Yang}},
  \bibinfo{journal}{Phys. Rev. Lett.} \textbf{\bibinfo{volume}{89}},
  \bibinfo{pages}{180601} (\bibinfo{year}{2002}).

\bibitem[{\citenamefont{Xiong}(2017)}]{xiong2017}
\bibinfo{author}{\bibfnamefont{D.}~\bibnamefont{Xiong}},
  \bibinfo{journal}{Phys. Rev. E} \textbf{\bibinfo{volume}{95}},
  \bibinfo{pages}{062140} (\bibinfo{year}{2017}).

\bibitem[{\citenamefont{Li and Li}(2018)}]{Li2018}
\bibinfo{author}{\bibfnamefont{S.}~\bibnamefont{Li}} \bibnamefont{and}
  \bibinfo{author}{\bibfnamefont{N.}~\bibnamefont{Li}}, \bibinfo{journal}{Sci.
  Rep.} \textbf{\bibinfo{volume}{8}}, \bibinfo{pages}{5329}
  (\bibinfo{year}{2018}), ISSN \bibinfo{issn}{{2045-2322}}.

\bibitem[{\citenamefont{Deutsch and Narayan}(2003)}]{Deutsch2003}
\bibinfo{author}{\bibfnamefont{J.~M.} \bibnamefont{Deutsch}} \bibnamefont{and}
  \bibinfo{author}{\bibfnamefont{O.}~\bibnamefont{Narayan}},
  \bibinfo{journal}{Phys. Rev. E} \textbf{\bibinfo{volume}{68}},
  \bibinfo{pages}{010201} (\bibinfo{year}{2003}).

\bibitem[{\citenamefont{Lepri}(1998)}]{Lepri1998a}
\bibinfo{author}{\bibfnamefont{S.}~\bibnamefont{Lepri}},
  \bibinfo{journal}{Phys. Rev. E} \textbf{\bibinfo{volume}{58}},
  \bibinfo{pages}{7165} (\bibinfo{year}{1998}).

\bibitem[{\citenamefont{Narayan and Ramaswamy}(2002)}]{Narayan2002b}
\bibinfo{author}{\bibfnamefont{O.}~\bibnamefont{Narayan}} \bibnamefont{and}
  \bibinfo{author}{\bibfnamefont{S.}~\bibnamefont{Ramaswamy}},
  \bibinfo{journal}{Phys. Rev. Lett.} \textbf{\bibinfo{volume}{89}},
  \bibinfo{pages}{200601} (\bibinfo{year}{2002}).

\bibitem[{\citenamefont{van Beijeren}(2012)}]{Beijeren2012}
\bibinfo{author}{\bibfnamefont{H.}~\bibnamefont{van Beijeren}},
  \bibinfo{journal}{Phys. Rev. Lett.} \textbf{\bibinfo{volume}{108}},
  \bibinfo{pages}{180601} (\bibinfo{year}{2012}).

\bibitem[{\citenamefont{Spohn}(2014)}]{Spohn2014}
\bibinfo{author}{\bibfnamefont{H.}~\bibnamefont{Spohn}}, \bibinfo{journal}{J
  Stat. Phys.} \textbf{\bibinfo{volume}{154}}, \bibinfo{pages}{1191}
  (\bibinfo{year}{2014}).

\bibitem[{\citenamefont{Tsironis et~al.}(1999)\citenamefont{Tsironis, Bishop,
  Savin, and Zolotaryuk}}]{Tsironis1999}
\bibinfo{author}{\bibfnamefont{G.~P.} \bibnamefont{Tsironis}},
  \bibinfo{author}{\bibfnamefont{A.~R.} \bibnamefont{Bishop}},
  \bibinfo{author}{\bibfnamefont{A.~V.} \bibnamefont{Savin}}, \bibnamefont{and}
  \bibinfo{author}{\bibfnamefont{A.~V.} \bibnamefont{Zolotaryuk}},
  \bibinfo{journal}{Phys. Rev. E} \textbf{\bibinfo{volume}{60}},
  \bibinfo{pages}{6610} (\bibinfo{year}{1999}).

\bibitem[{\citenamefont{Hu et~al.}(2000)\citenamefont{Hu, Li, and
  Zhao}}]{Hu2000a}
\bibinfo{author}{\bibfnamefont{B.~B.} \bibnamefont{Hu}},
  \bibinfo{author}{\bibfnamefont{B.~W.} \bibnamefont{Li}}, \bibnamefont{and}
  \bibinfo{author}{\bibfnamefont{H.}~\bibnamefont{Zhao}},
  \bibinfo{journal}{Phys. Rev. E} \textbf{\bibinfo{volume}{61}},
  \bibinfo{pages}{3828} (\bibinfo{year}{2000}).

\bibitem[{\citenamefont{Aoki and Kusnezov}(2000)}]{Aoki2000}
\bibinfo{author}{\bibfnamefont{K.}~\bibnamefont{Aoki}} \bibnamefont{and}
  \bibinfo{author}{\bibfnamefont{D.}~\bibnamefont{Kusnezov}},
  \bibinfo{journal}{Phys Lett A} \textbf{\bibinfo{volume}{265}},
  \bibinfo{pages}{250} (\bibinfo{year}{2000}).

\bibitem[{\citenamefont{Prosen and Campbell}(2000)}]{Prosen2000}
\bibinfo{author}{\bibfnamefont{T.}~\bibnamefont{Prosen}} \bibnamefont{and}
  \bibinfo{author}{\bibfnamefont{D.~K.} \bibnamefont{Campbell}},
  \bibinfo{journal}{Phys. Rev. Lett.} \textbf{\bibinfo{volume}{84}},
  \bibinfo{pages}{2857} (\bibinfo{year}{2000}).

\bibitem[{\citenamefont{Giardina et~al.}(2000)\citenamefont{Giardina, Livi,
  Politi, and Vassalli}}]{Giardina2000}
\bibinfo{author}{\bibfnamefont{C.}~\bibnamefont{Giardina}},
  \bibinfo{author}{\bibfnamefont{R.}~\bibnamefont{Livi}},
  \bibinfo{author}{\bibfnamefont{A.}~\bibnamefont{Politi}}, \bibnamefont{and}
  \bibinfo{author}{\bibfnamefont{M.}~\bibnamefont{Vassalli}},
  \bibinfo{journal}{Phys. Rev. Lett.} \textbf{\bibinfo{volume}{84}},
  \bibinfo{pages}{2144} (\bibinfo{year}{2000}).

\bibitem[{\citenamefont{Li et~al.}(2015)\citenamefont{Li, Liu, Li, Hänggi, and
  Li}}]{Li2015}
\bibinfo{author}{\bibfnamefont{Y.}~\bibnamefont{Li}},
  \bibinfo{author}{\bibfnamefont{S.}~\bibnamefont{Liu}},
  \bibinfo{author}{\bibfnamefont{N.}~\bibnamefont{Li}},
  \bibinfo{author}{\bibfnamefont{P.}~\bibnamefont{Hänggi}}, \bibnamefont{and}
  \bibinfo{author}{\bibfnamefont{B.}~\bibnamefont{Li}}, \bibinfo{journal}{New J
  Phys} \textbf{\bibinfo{volume}{17}}, \bibinfo{pages}{043064}
  (\bibinfo{year}{2015}).

\bibitem[{\citenamefont{Nika and Balandin}(2017)}]{Nika2017}
\bibinfo{author}{\bibfnamefont{D.~L.} \bibnamefont{Nika}} \bibnamefont{and}
  \bibinfo{author}{\bibfnamefont{A.~A.} \bibnamefont{Balandin}},
  \bibinfo{journal}{Rep. Prog. Phys.} \textbf{\bibinfo{volume}{80}},
  \bibinfo{pages}{036502} (\bibinfo{year}{2017}).

\bibitem[{\citenamefont{Lippi and Livi}(2000)}]{Lippi2000a}
\bibinfo{author}{\bibfnamefont{A.}~\bibnamefont{Lippi}} \bibnamefont{and}
  \bibinfo{author}{\bibfnamefont{R.}~\bibnamefont{Livi}}, \bibinfo{journal}{J.
  Stat. Phys.} \textbf{\bibinfo{volume}{100}}, \bibinfo{pages}{1147}
  (\bibinfo{year}{2000}).

\bibitem[{\citenamefont{Xiong et~al.}(2010)\citenamefont{Xiong, Wang, Zhang,
  and Zhao}}]{PhysRevE.82.030101}
\bibinfo{author}{\bibfnamefont{D.}~\bibnamefont{Xiong}},
  \bibinfo{author}{\bibfnamefont{J.}~\bibnamefont{Wang}},
  \bibinfo{author}{\bibfnamefont{Y.}~\bibnamefont{Zhang}}, \bibnamefont{and}
  \bibinfo{author}{\bibfnamefont{H.}~\bibnamefont{Zhao}},
  \bibinfo{journal}{Phys. Rev. E} \textbf{\bibinfo{volume}{82}},
  \bibinfo{pages}{030101} (\bibinfo{year}{2010}).

\bibitem[{\citenamefont{Shiba and Ito}(2008)}]{Shiba2008a}
\bibinfo{author}{\bibfnamefont{H.}~\bibnamefont{Shiba}} \bibnamefont{and}
  \bibinfo{author}{\bibfnamefont{N.}~\bibnamefont{Ito}}, \bibinfo{journal}{J
  Phys Soc Jpn} \textbf{\bibinfo{volume}{77}}, \bibinfo{pages}{054006}
  (\bibinfo{year}{2008}).

\bibitem[{\citenamefont{Wang et~al.}(2012)\citenamefont{Wang, Hu, and
  Li}}]{Wang2012a}
\bibinfo{author}{\bibfnamefont{L.}~\bibnamefont{Wang}},
  \bibinfo{author}{\bibfnamefont{B.}~\bibnamefont{Hu}}, \bibnamefont{and}
  \bibinfo{author}{\bibfnamefont{B.}~\bibnamefont{Li}}, \bibinfo{journal}{Phys.
  Rev. E} \textbf{\bibinfo{volume}{86}}, \bibinfo{pages}{040101}
  (\bibinfo{year}{2012}).

\bibitem[{\citenamefont{Savin et~al.}(2016)\citenamefont{Savin, Zolotarevskiy,
  and Gendelman}}]{Savin2016}
\bibinfo{author}{\bibfnamefont{A.~V.} \bibnamefont{Savin}},
  \bibinfo{author}{\bibfnamefont{V.}~\bibnamefont{Zolotarevskiy}},
  \bibnamefont{and} \bibinfo{author}{\bibfnamefont{O.~V.}
  \bibnamefont{Gendelman}}, \bibinfo{journal}{Europhys. Lett.}
  \textbf{\bibinfo{volume}{113}}, \bibinfo{pages}{24003}
  (\bibinfo{year}{2016}).

\bibitem[{\citenamefont{Liu et~al.}(2014)\citenamefont{Liu, Haenggi, Li, Ren,
  and Li}}]{Liu2014b}
\bibinfo{author}{\bibfnamefont{S.}~\bibnamefont{Liu}},
  \bibinfo{author}{\bibfnamefont{P.}~\bibnamefont{Haenggi}},
  \bibinfo{author}{\bibfnamefont{N.}~\bibnamefont{Li}},
  \bibinfo{author}{\bibfnamefont{J.}~\bibnamefont{Ren}}, \bibnamefont{and}
  \bibinfo{author}{\bibfnamefont{B.}~\bibnamefont{Li}}, \bibinfo{journal}{Phys.
  Rev. Lett.} \textbf{\bibinfo{volume}{112}}, \bibinfo{pages}{040601}
  (\bibinfo{year}{2014}).

\bibitem[{\citenamefont{Cheng et~al.}(2014)\citenamefont{Cheng, Grossman, and
  McKercher}}]{cuda}
\bibinfo{author}{\bibfnamefont{J.}~\bibnamefont{Cheng}},
  \bibinfo{author}{\bibfnamefont{M.}~\bibnamefont{Grossman}}, \bibnamefont{and}
  \bibinfo{author}{\bibfnamefont{T.}~\bibnamefont{McKercher}},
  \emph{\bibinfo{title}{Professional Cuda C Programming}}
  (\bibinfo{publisher}{John Wiley \& Sons, Inc.}, \bibinfo{year}{2014}).

\bibitem[{\citenamefont{Zhao}(2006)}]{H.Zhao2006}
\bibinfo{author}{\bibfnamefont{H.}~\bibnamefont{Zhao}}, \bibinfo{journal}{Phys.
  Rev. Lett.} \textbf{\bibinfo{volume}{96}}, \bibinfo{pages}{140602}
  (\bibinfo{year}{2006}).

\bibitem[{\citenamefont{Chen et~al.}(2013)\citenamefont{Chen, Zhang, Wang, and
  Zhao}}]{Chen2013}
\bibinfo{author}{\bibfnamefont{S.}~\bibnamefont{Chen}},
  \bibinfo{author}{\bibfnamefont{Y.}~\bibnamefont{Zhang}},
  \bibinfo{author}{\bibfnamefont{J.}~\bibnamefont{Wang}}, \bibnamefont{and}
  \bibinfo{author}{\bibfnamefont{H.}~\bibnamefont{Zhao}},
  \bibinfo{journal}{Phys. Rev. E} \textbf{\bibinfo{volume}{87}},
  \bibinfo{pages}{032153} (\bibinfo{year}{2013}).

\bibitem[{\citenamefont{Gao et~al.}(2016)\citenamefont{Gao, Li, and
  Li}}]{Gao2016c}
\bibinfo{author}{\bibfnamefont{Z.}~\bibnamefont{Gao}},
  \bibinfo{author}{\bibfnamefont{N.}~\bibnamefont{Li}}, \bibnamefont{and}
  \bibinfo{author}{\bibfnamefont{B.}~\bibnamefont{Li}}, \bibinfo{journal}{Phys.
  Rev. E} \textbf{\bibinfo{volume}{93}}, \bibinfo{pages}{022102}
  (\bibinfo{year}{2016}), ISSN \bibinfo{issn}{2470-0045}.

\bibitem[{\citenamefont{Xiong}(2018)}]{Xiong2018a}
\bibinfo{author}{\bibfnamefont{D.}~\bibnamefont{Xiong}},
  \bibinfo{journal}{Phys. Rev. E} \textbf{\bibinfo{volume}{97}},
  \bibinfo{pages}{022116} (\bibinfo{year}{2018}).

\bibitem[{\citenamefont{Xiong and Zhang}(2018)}]{Xiong2018b}
\bibinfo{author}{\bibfnamefont{D.}~\bibnamefont{Xiong}} \bibnamefont{and}
  \bibinfo{author}{\bibfnamefont{Y.}~\bibnamefont{Zhang}},
  \bibinfo{journal}{Phys. Rev. E} \textbf{\bibinfo{volume}{98}},
  \bibinfo{pages}{012130} (\bibinfo{year}{2018}).

\bibitem[{\citenamefont{Yang et~al.}(2006)\citenamefont{Yang, Grassberger, and
  Hu}}]{Yang2006a}
\bibinfo{author}{\bibfnamefont{L.}~\bibnamefont{Yang}},
  \bibinfo{author}{\bibfnamefont{P.}~\bibnamefont{Grassberger}},
  \bibnamefont{and} \bibinfo{author}{\bibfnamefont{B.}~\bibnamefont{Hu}},
  \bibinfo{journal}{Phys. Rev. E} \textbf{\bibinfo{volume}{74}},
  \bibinfo{pages}{062101} (\bibinfo{year}{2006}).

\bibitem[{\citenamefont{Xiong and Dmitriev}(2019)}]{phi4book}
\bibinfo{author}{\bibfnamefont{D.-X.} \bibnamefont{Xiong}} \bibnamefont{and}
  \bibinfo{author}{\bibfnamefont{S.-V.} \bibnamefont{Dmitriev}},
  \emph{\bibinfo{title}{A Dynamical Perspective on the {$\phi^4$} Model (Pages
  281-308)}} (\bibinfo{publisher}{Springer, Cham}, \bibinfo{year}{2019}).

\bibitem[{\citenamefont{Das et~al.}(2014)\citenamefont{Das, Dhar, Saito, Mendl,
  and Spohn}}]{Das2014a}
\bibinfo{author}{\bibfnamefont{S.~G.} \bibnamefont{Das}},
  \bibinfo{author}{\bibfnamefont{A.}~\bibnamefont{Dhar}},
  \bibinfo{author}{\bibfnamefont{K.}~\bibnamefont{Saito}},
  \bibinfo{author}{\bibfnamefont{C.~B.} \bibnamefont{Mendl}}, \bibnamefont{and}
  \bibinfo{author}{\bibfnamefont{H.}~\bibnamefont{Spohn}},
  \bibinfo{journal}{Phys. Rev. E} \textbf{\bibinfo{volume}{90}},
  \bibinfo{pages}{012124} (\bibinfo{year}{2014}).

\bibitem[{\citenamefont{Lee-Dadswell et~al.}(2005)\citenamefont{Lee-Dadswell,
  Nickel, and Gray}}]{Lee-Dadswell2005}
\bibinfo{author}{\bibfnamefont{G.~R.} \bibnamefont{Lee-Dadswell}},
  \bibinfo{author}{\bibfnamefont{B.~G.} \bibnamefont{Nickel}},
  \bibnamefont{and} \bibinfo{author}{\bibfnamefont{C.~G.} \bibnamefont{Gray}},
  \bibinfo{journal}{Phys. Rev. E} \textbf{\bibinfo{volume}{72}},
  \bibinfo{pages}{031202} (\bibinfo{year}{2005}).

\bibitem[{\citenamefont{Zaburdaev et~al.}(2015)\citenamefont{Zaburdaev,
  Denisov, and Klafter}}]{Zaburdaev2015}
\bibinfo{author}{\bibfnamefont{V.}~\bibnamefont{Zaburdaev}},
  \bibinfo{author}{\bibfnamefont{S.}~\bibnamefont{Denisov}}, \bibnamefont{and}
  \bibinfo{author}{\bibfnamefont{J.}~\bibnamefont{Klafter}},
  \bibinfo{journal}{Rev Mod Phys} \textbf{\bibinfo{volume}{87}},
  \bibinfo{pages}{483} (\bibinfo{year}{2015}).

\bibitem[{\citenamefont{Klafter and Sokolov}(2011)}]{firststeprw}
\bibinfo{author}{\bibfnamefont{J.}~\bibnamefont{Klafter}} \bibnamefont{and}
  \bibinfo{author}{\bibfnamefont{I.~M.} \bibnamefont{Sokolov}},
  \emph{\bibinfo{title}{First Steps in Random Walks: From Tools to
  Applications}} (\bibinfo{publisher}{OUP Oxford}, \bibinfo{year}{2011}).

\end{thebibliography}

\end{document}